\documentclass[]{jfm}

\usepackage{graphicx}
\usepackage{newtxtext}
\usepackage{newtxmath}
\usepackage{natbib}
\usepackage{hyperref}
\hypersetup{
    colorlinks = true,
    urlcolor   = blue,
    citecolor  = black,
}

\newcommand{\RomanNumeralCaps}[1]
\linenumbers

\newcommand\Pra{\mbox{\textit{Pr}}}
\newcommand\Sch{\mbox{\textit{Sc}}}
\newcommand\Ray{\mbox{\textit{Ra}}}

\newcommand\She{\mbox{\textit{Sh}}}
\newcommand\Dar{\mbox{\textit{Da}}}

\makeatletter
\newcommand*{\rom}[1]{\expandafter\@slowromancap\romannumeral #1@}
\makeatother

\title{From Darcy flow to convective flow: pore-scale study of density-driven currents in porous media}

\author{Junyi Li\aff{1},
	Yantao Yang\aff{2,3}\corresp{\email{yantao.yang@pku.edu.cn}}
	\and Chao Sun\aff{1,4}\corresp{\email{chaosun@tsinghua.edu.cn}}}

\affiliation{
	\aff{1} New Cornerstone Science Laboratory, Center for Combustion Energy, Key Laboratory for
	Thermal Science and Power Engineering of Ministry of Education, Department of Energy and
	Power Engineering, Tsinghua University, Beijing 100871, P. R. China  \\
	\aff{2} State Key Laboratory for Turbulence and Complex Systems, and Department of Mechanics and Engineering Science, College of Engineering, Peking University, Beijing 100871, P. R. China \\
	\aff{3} Joint Laboratory of Marine Hydrodynamics and Ocean Engineering, Laoshan Laboratory, Shandong 266299, P. R. China \\
	\aff{4} Department of Engineering Mechanics, School of Aerospace Engineering, Tsinghua University, Beijing 100871, P. R. China
}

\begin{document}
\maketitle

\begin{abstract}
We conducted a series of pore-scale numerical simulations on convective flow in porous media, with a fixed Schmidt number of 400 and a wide range of Rayleigh numbers. The porous media are modeled using regularly arranged square obstacles in a Rayleigh-Bénard (RB) system. As the Rayleigh number increases, the flow transitions from a Darcy-type regime to an RB-type regime, with the corresponding $Sh-Ra_D$ relationship shifting from sublinear scaling to the classical 0.3 scaling of RB convection. Here, $Sh$ and $Ra_D$ represent the Sherwood number and Rayleigh-Darcy number, respectively. For different porosities, the transition begins at approximately $Ra_D = 4000$, at which point the characteristic horizontal scale of the flow field is comparable to the size of a single obstacle unit. When the thickness of the concentration boundary layer is less than about one-sixth of the pore spacing, the flow finally enters the RB regime. In the Darcy regime, the scaling exponent of $Sh$ and $Ra_D$ decreases as porosity increases. Based on the Grossman-Lohse theory (\textit{J. Fluid Mech}., vol. 407, 2000; \textit{Phys. Rev. Lett}., vol. 86, 2001), we provide an explanation for the scaling laws in each regime and highlight the significant impact of mechanical dispersion effects within the boundary layer. Our findings provide some new insights into the validity range of the Darcy model.
\end{abstract}


\section{Introduction}\label{sec:intro}
Porous media refer to materials composed of numerous frameworks that create many tiny voids. When a fluid fills these voids, convection may occur under the influence of gravity due to density differences. This density-driven convection in porous media exhibits properties that differ from those in free fluid convection and is widely present in nature and engineering applications, such as the formation of sea ice and the oil recovery from geological formations \citep{farajzadeh2012foam,miah2018,anderson2022}. In recent years, due to the rise of carbon dioxide (CO$_2$) geological sequestration, convection in porous media has become a research hotspot \citep{huppert2014fluid,bachu2015,de2021}. This technology involves capturing CO$_2$ generated from industrial processes and burying it underground to reduce carbon emissions. When CO$_2$ is injected into deep saline aquifers, it enters a supercritical state similar to a fluid and initially rises to an impermeable layer due to its lower density. Subsequently, CO$_2$ diffuses horizontally and gradually dissolves in the underlying saline water. This dissolved liquid is denser than the surrounding fluid, thus driving convection. The dissolution and convection process of CO$_2$ in saline aquifers is crucial for preventing leakage and ensuring its long-term storage. Therefore, understanding the mass transport efficiency in porous media convection has become a key focus of research \citep{hewitt2020vigorous,de2023convective}.

The commonly used model for convective flow in porous media is based on Darcy's law (hereafter referred to as the Darcy model) \citep{nield2017}. The smallest analytical flow domain in the Darcy model includes multiple pores and is known as the Representative Elementary Volume (REV). The scale of the REV is generally much larger than the pore scale but much smaller than the characteristic flow scale. Consequently, the Darcy model solves for macroscopic variables that are volume-averaged, neglecting information at the pore scale. In this model, the only governing parameter is the Rayleigh-Darcy number $\Ray_D$. Early theoretical studies suggest that for asymptotically large $\Ray_D$, the Sherwood number $\She$, which represents the efficiency of mass transport, scales linearly with $\Ray_D$ \citep{malkus1954,howard1966convection}. With the improvement of computational capabilities in recent years, numerous numerical simulations for high-$Ra_D$ Darcy model have emerged \citep{hewitt2012ultimate,hewitt2013convective,hewitt2014high,pirozzoli2021towards,de2022strong}. The results of two-dimensional (2D) simulations confirm the linear scaling of $\She$ versus $\Ray_D$. However, three-dimensional (3D) simulations have revealed additional sublinear term. This discrepancy is primarily due to the fact that in 3D simulations, $\Ray_D$ has not reached a sufficiently high value for the flow to enter the ultimate regime \citep{de2023convective}. Recently, \citet{zhu2024} successfully utilized the Grossman-Lohse (GL) theory \citep{grossmann2000scaling,grossmann2001thermal} to explain the differences between the 2D and 3D results. 

Although numerical simulations align with the theory, numerous laboratory experiments have consistently found that the scaling exponent of $Sh$ versus $Ra_D$ is always less than 1 \citep{neufeld2010convective,backhaus2011convective,wang2016three,liang2018effect}. This indicates that pore-scale effects cannot be neglected when considering mass transport in convective flows through actual porous media. To incorporate these effects into simulations, there are generally two approaches. One is to introduce additional terms into the Darcy model to account for the interactions between solid structures and the fluids. For example, when the flow velocity within the pores is relatively high, the drag exerted by solid obstacles on the flow becomes non-negligible, requiring the consideration of the Forchheimer term \citep{joseph1982nonlinear,nield2017,jin2024}. In recent years, there has been increased attention on mechanical dispersion, which refer to the alteration of flow direction and further mixing of solutes due to pore structures \citep{saffman1959theory,dentz2023mixing}. This dispersion effect produces results similar to those caused by molecular diffusion; therefore, the two are often considered together under the term hydrodynamic dispersion \citep{de2023convective}. In the Darcy simulations, to account for the effects of dispersion, the molecular diffusion coefficient is typically replaced by the Fickian dispersion tensor \citep{bear1961tensor,hidalgo2009effect,ghesmat2011effect}. Recent simulation \citep{wen2018rayleigh} reported a fan-flow structure caused by dispersion effect, which indeed influence mass transport. However, because these models introduce numerous parameters, extensive experiments and numerical simulations are still required to calibrate and validate them.

Another approach involves conducting pore-scale simulations, which solves the original Navier-Stokes (NS) equations within the pores, rather than using the Darcy model. This places extremely high demands on computational capabilities, particularly for cases with low porosity. In our previous work \citep{liu2020rayleigh,liu2021lagrangian}, we conducted 2D pore-scale simulations on thermal convection with relatively high porosities and found that the heat transfer first increases and then decreases as porosity decreases. 
This is due to the fact that obstacles simultaneously hinder convective flow and improve the continuity of the flow structure. Recently, \citet{Gasow2020,gasow2021,gasow2022} conducted a series of 2D pore-scale simulations at large Schmidt numbers with porosities as low as 0.09, discovering that the scaling exponent of $Sh$ versus $Ra_D$ decreases with increasing porosity, which aligns with experimental results. Moreover, they pointed out that the discrepancy between pore-scale simulations and Darcy simulations might be due to the latter's omission of the momentum dispersion term.

One concern about pore-scale simulations is that as the control parameter, i.e. the Rayleigh number $\Ray$, increases, the characteristic scale of the flow field gradually decreases, and the flow may eventually transition into a state similar to Rayleigh-Bénard (RB) convection that is no longer influenced by the porous media. Both our previous work \citep{liu2020rayleigh} and experiments \citep{ataei2019} have shown that under extremely high $\Ray$, the heat transport efficiency follows the classical 0.3 scaling of RB convection \citep{grossmann2000scaling,grossmann2001thermal}. It is important to note that in this type of thermal convection, obstacles can conduct heat, whereas for solutes, they cannot pass through the obstacles. For the latter case, would the flow also eventually transition to an RB state? \citet{Gasow2020,gasow2021,gasow2022}'s simulations did not reveal any clear indications of this, possibly because the control parameter was not large enough. In this paper, we conduct systematic 2D pore-scale simulations over a wide range of $\Ray$ to investigate this issue, aiming to illustrate when the flow transitions from a Darcy-type to an RB-type regime and to thoroughly examine the characteristics of the transitional phase. We also aim to utilize the GL theory to analyze the transport scaling laws in different regimes. It should be noted that our pore-scale analysis fundamentally differs from the work of \citet{zhu2024}, as their study was conducted on the macroscopic (Darcy) scale.

This paper is organized as follows. In section \ref{sec:form} we introduce the governing equations and the numerical settings. Then in section \ref{sec:structure} and section \ref{sec:transfer} we present the numerical results, including the variations of flow structures and mass transfer, respectively. In section \ref{sec:GL} we provide a theoretically analysis based on the GL theory. Finally, we give the conclusions in section \ref{sec:conclusion}.
\section{Problem formulation}\label{sec:form}

\subsection{Governing equations}
\begin{figure}
 \centering
 \includegraphics[width=0.8\textwidth]{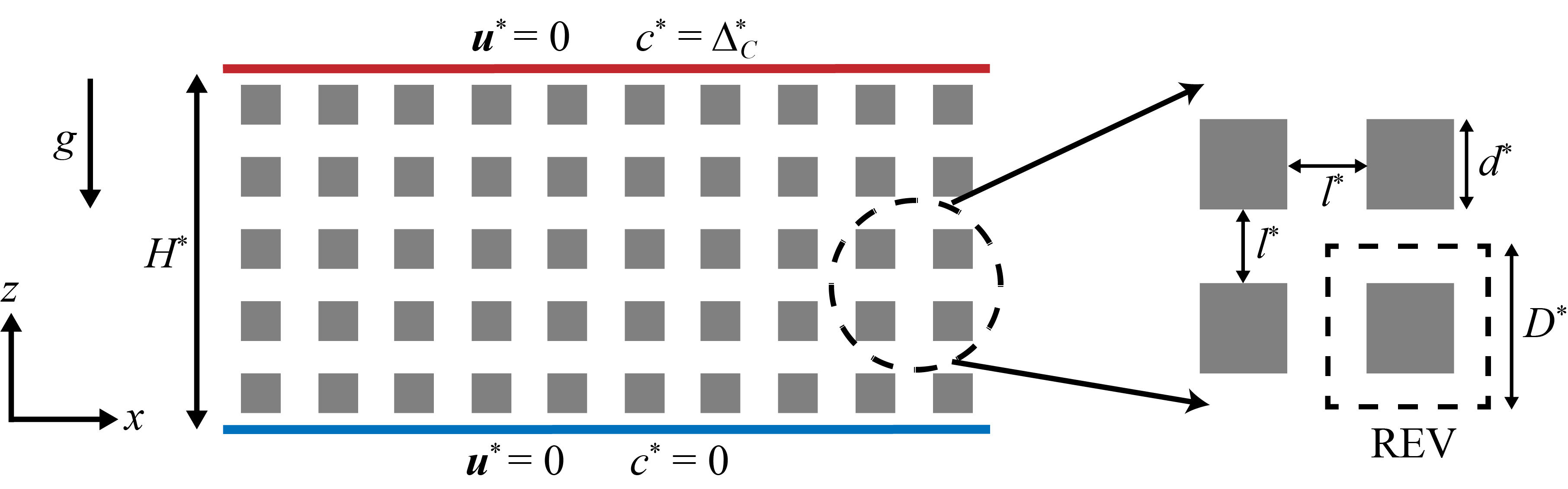}%
 \caption{Schematic illustration of the two-dimensional flow domain. The regular porous media is represented by the grey obstacles. The size of a REV is $D^*=d^*+l^*$.}
 \label{fig:sketch}
\end{figure}
We consider a 2D RB system filled with regular square obstacles, as shown in figure \ref{fig:sketch}. Constant species concentrations are kept on the two horizontal plates with a separation of $H^*$. Hereafter, the asterisk $*$ represents the dimensional forms of the variables. $c^*$ is the relative species concentration, which equals zero at the bottom plate and $\Delta_c^*$ at the top plate. The fluid density is then determined by $c^*$ as $\rho^*=\rho_0^*(1+\beta c^*)$, in which $\rho_0^*$ is the density of a reference state, and $\beta$ is the concentration expansion coefficient. The periodic boundary condition is used in the horizontal direction of the flow domain. The no-slip boundary condition is applied at all solid surfaces, including the two plates and the fluid-obstacle interfaces. In addition. the obstacles are assumed to be non-penetrated by the species. Within the Oberbeck-Boussinesq approximation, the incompressible flow in the pores is governed by
\begin{subequations}\label{eq:ns*}
\begin{eqnarray}
  	\frac{\partial \boldsymbol{u}^*}{\partial t^*}  + \boldsymbol{u}^*\bcdot\bnabla\boldsymbol{u}^* &=& -\bnabla p^* + \nu \bnabla^2\boldsymbol{u}^* - g \beta c^*\boldsymbol{e}_z,  \\
	\frac{\partial c^*}{\partial t^*}+ \boldsymbol{u}^*\bcdot\bnabla c^* &=& \kappa \nabla^2 c^*, \label{eq:ns*b}\\
	\bnabla\bcdot\boldsymbol{u}^* &=& 0. 
\end{eqnarray}
\end{subequations} 
Here, $\boldsymbol{u}^*$ is the velocity vector, $p^*$ is pressure, $\nu$ is viscosity, $g$ is the gravitational acceleration, $\boldsymbol{e}_z$ is the vertical unit vector, and $\kappa$ is the mass diffusivity of the species, respectively. We use the domain height $H^*$, the free-fall velocity $\sqrt{gH^*\beta\Delta_c^*}$, and the concentration difference $\Delta_c^*$ to non-dimensionalise the governing equations \eqref{eq:ns*}. Dimensionless control parameters include the aspect ratio, the Schmidt number and the Rayleigh number, which are defined respectively as
\begin{equation}
\Gamma = \frac{L^*}{H^*}, \quad
\Sch = \frac{\nu}{\kappa}, \quad
\Ray=\frac{g\beta \Delta_c^* {H^*}^3}{\kappa\nu}.
\end{equation}
Here, $L^*$ is the domain width. Then the dimensionless governing equations read
\begin{subequations}\label{eq:ns}
\begin{eqnarray}
  	\frac{\partial \boldsymbol{u}}{\partial t} + \boldsymbol{u}\bcdot\bnabla\boldsymbol{u} &=& -\bnabla p + \sqrt{\frac{\Sch}{\Ray}}\bnabla^2\boldsymbol{u} - c\boldsymbol{e}_z, \label{eq:ns_a} \\
	\frac{\partial c}{\partial t}+ \boldsymbol{u}\bcdot\bnabla c &=& \frac{1}{\sqrt{\Sch\Ray}} \nabla^2 c, \label{eq:ns_b} \\
	\bnabla\bcdot\boldsymbol{u} &=& 0, 
\end{eqnarray}
\end{subequations} 
and the boundary conditions at two plates read
\begin{subequations} \label{eq:bc}
	\begin{eqnarray}
	\boldsymbol{u}&=&0, \quad c=0,\qquad at \quad z=0; \\
	\boldsymbol{u}&=&0, \quad c=1,\qquad at \quad z=1.
	\end{eqnarray}
\end{subequations} 
The key response parameters include the Sherwood number and the Reynolds number, which are defined respectively as
\begin{equation}
 \She = \frac{\kappa \langle \frac{\partial c^*}{\partial z^*}\rangle_{z^*=0}  }{\kappa \Delta^*_c {H^*}^{-1}}=\langle \frac{\partial c}{\partial z}\rangle_{z=0} , \quad
 \Rey = \frac{\boldsymbol{u}^*_{rms} H^*}{\nu}=\sqrt{\frac{\Ray}{\Sch}}\boldsymbol{u}_{rms}. \label{eq:response}
\end{equation}
Hereafter, the bracket $\langle \cdot \rangle$ denotes the average over a specific horizontal plane. The subscript `rms' stands for the root-mean-square value over the entire domain except for obstacles. Note that the Sherwood number $\She$ represents the ratio of the total mass transfer rate to the rate of diffusive mass transport. In the statistically steady state, $\She$ should be the same at the two plates. 

\subsection{Porous media} 
\begin{table} 
\begin{center}
\def~{\hphantom{0}}
\begin{tabular}{cccccccc}
$\phi$ & $\Gamma$  & $N_{px}$ & $N_{pz}$ & $d$ & $l$ & $D$ & $Da$ \\[5pt]
0.64 &  2   &  30  &  15  & 0.04 & 0.0267  &  0.0667  &  $2.6 \times 10^{-5}$\\ 
0.36 &  2   &  40  &  20  & 0.04 & 0.0100  &  0.0500  &  $1.4 \times 10^{-6}$\\ 
0.15 &  1   &  23  &  23  & 0.04 & 0.0035  &  0.0435  & $6.4\times10^{-8}$\\ 
\end{tabular}
\caption{The details for the porous media. Columns from left to right are: the porosity, the aspect ratio, the number of obstacles in the horizontal direction and the vertical direction, the size and the spacing of the obstacles, the size of the REVs and the Darcy number.}
\label{tab:porous}
\end{center}
\end{table}
The porous medium structures set in this study are all regularly arranged squares, as shown in figure \ref{fig:sketch}. The side length of each square is $d^* (=dH^*)$, and the spacing between them is $l^* (=lH^*)$. In traditional macroscopic simulations of convection in porous media, the REV which contains several pores is generally defined as the basic research unit \citep{nield2017}. In our pore-scale simulations, we can similarly define a square REV, whose side length is $D^*=d^*+l^*$. Under this definition, the whole domain is filled with REVs. The additional control parameters related to the porous medium structure include the porosity, the Darcy number and the Rayleigh-Darcy number, which are defined as follows:
\begin{equation}  \label{eq:porous}
 \phi=1-(\frac{d^*}{D^*})^2,\quad \Dar=\frac{K^*}{{H^*}^2}, \quad \Ray_D=\frac{g\beta \Delta_c^* H^* K^*}{\kappa\nu}=\Ray \Dar.
\end{equation}
Here, $K^*$ is the permeability, which can be estimated by the Kozeny's equation \citep{nield2017}:
\begin{equation}  \label{eq:K}
K^*=\frac{{d^*}^2\phi^3}{\eta(1-\phi)^2},
\end{equation}
in which $\eta$ is an empirical coefficient. In this study, we use $\eta=126$, a value derived from the fitting coefficient in \citet{Gasow2020}'s simulations, where their porous medium configuration also consists of regularly arranged squares. Table \ref{tab:porous} summarizes the three sets of porous medium configurations we have defined, with porosity of 0.64, 0.36, and 0.15, respectively. In these configurations, we keep the size of the obstacles constant and vary their number to change the porosity. Accordingly, the spacing between the obstacles and the size of the REV also change. The Darcy number is calculated by substitute \eqref{eq:K} into \eqref{eq:porous}. It should be noted that in actual CO$_2$ sequestration, the Darcy number is very small, typically reaching the order of $10^{-13}$ \citep{jin2024}. For pore-scale simulations, achieving such small pore sizes would require very high resolution, which is computationally infeasible with current capabilities. In the series of works by \citet{Gasow2020,gasow2021,gasow2022}, the lowest Darcy number achieved was on the order of $10^{-8}$, the same as in this study. Nevertheless, the flow behaviors observed at larger Darcy numbers show a certain consistency, allowing for reasonable extrapolation to cases with smaller Darcy numbers.

In the numerical simulations of this study, the boundary conditions of the obstacles are implemented using the direct-forcing immersed boundary method (IBM) \citep{uhlmann2005}. Specifically, an additional body force $\boldsymbol{f}$ is introduced in the momentum equation \eqref{eq:ns_a} and the convection-diffusion equation \eqref{eq:ns_b} to ensure zero velocity and zero normal gradient of concentration at the obstacle boundaries. Particularly, at the corners of the square, the sum of the fluxes of the concentration field along the $x$ and $z$ directions is set to zero. Under such settings, the flow state inside the obstacles does not affect the external flow field. This method has been widely used in various complex geometries and deformable interfaces \citep{spandan2017,Vanella2020}, and was also employed in our previous simulations involving circular obstacles \citep{liu2020rayleigh}. For the current square obstacles, their boundaries are directly defined on the Eulerian grid, thus avoiding additional errors due to interpolation. To ensure accuracy, we extend two grid layers inward from the obstacle boundaries, enforcing both no-slip and no-penetration boundary conditions. Additionally, we have conducted some a posteriori validations, with details provided in the next section.

In the current simulations, the obstacles are arranged in a regular pattern, which differs from the real irregular porous structures found in geological formations. However, it is extremely challenging to set up randomly distributed obstacles in pore-scale simulations, especially in cases with very low porosity. The difficulty mainly lies in the narrow spacing between obstacles caused by randomness and the complex structures formed by the combination of multiple obstacles. These require highly precise grids for resolution, which poses a significant challenge given the current computational capabilities. Therefore, we continue to focus on cases with regularly arranged obstacles. We believe that the conclusions drawn from this basic configuration can provide us with a deeper understanding of the physical mechanisms of convection in porous media. Additionally, we conducted a separate simulation with randomly distributed obstacles for a porosity of 0.64, and the results are summarized in the appendix \ref{appA}. It can be observed that the qualitative trend of mass transport is consistent between the regular and random distributions of obstacles, although there are some minor quantitative differences. A systematic study on this aspect is still needed in future research.

\subsection{Numerical settings} 

\begin{table} 
\begin{center}
\def~{\hphantom{0}}
\begin{tabular}{cccccccccccccccc}
  $\phi$ & $\Ray$ & $\Ray_D$ & $\Gamma$  & $N_x(m_x)$ & $N_z(m_z)$ & $t_d$ & $t_s$ & $\overline{\She}$ & $\overline{\Rey}$ & $\Delta_{Sh}(\%)$\\[5pt]
   0.64 & $1 \times 10^7$ & $2.6 \times 10^2$ &  2   &   512(2)  &   256(2)  &  20000 &  5000  &  2.307  & 0.119 & 0.4 \\ 
   0.64 & $4 \times 10^7$ & $1.0 \times 10^3$ &  2   &   512(2)  &   256(2)  &  10000 &  5000  &  8.344  & 0.470 & 0.7  \\ 
    0.64 & $6 \times 10^7$ & $1.5 \times 10^3$ &  2   &   512(2)  &   256(2)  &  10000 &  5000  &  11.71  & 0.685 & 2.9  \\ 
   0.64 & $1 \times 10^8$ & $2.6 \times 10^3$ &  2   &   768(2)  &   384(2)  &  10000 &  40000  &  18.17  & 1.094  &  0.4  \\ 
   0.64 & $1.4 \times 10^8$ & $3.6 \times 10^3$ &  2   &   768(2)  &   384(2)  &  10000 &  50000  &  23.17  & 1.461  &  0.3  \\ 
   0.64 & $2\times 10^8$  & $5.1 \times 10^3$ &  2   &   768(2)  &   384(2)  &  5000 &  2000  &  29.51  & 1.964  &  2.2  \\ 
   0.64 & $4 \times 10^8$ & $1.0 \times 10^4$ &  2   &   768(2)  &   384(2)  &  5000 &  2000  &  42.32  & 3.323  &  0.8  \\ 
   0.64 & $1 \times 10^9$ & $2.6 \times 10^4$ &  2   &   1152(3)  &   576(3)  &  2000  &  2000  & 63.25 &  6.480  &  0.5  \\ 
   0.64 & $2 \times 10^9$ & $5.1 \times 10^4$ &  2   &   1152(3)  &   576(3)  &  2000  &  1000  & 75.16 &  9.840  &  0.5 \\ 
   0.64 & $4 \times 10^9$ & $1.0 \times 10^5$ &  2   &   1152(3)  &   576(3)  &  2000  &  1000  & 94.37 &  15.62  &  0.4  \\ 
   0.64 & $1 \times 10^{10}$ & $2.6 \times 10^5$ &  2  &   1280(4)  &   640(4)  &   1000  &  1000 &  133.3  &  29.29  &  0.3 \\ 
   0.64 & $1 \times 10^{11}$ & $2.6 \times 10^6$ &  2  &  1536(6)  &   768(6)  &  1000  &  1000  &  249.1  &  126.5  & 0.04 \\  [5pt]
			
  0.36 &$1 \times 10^8$ &$1.4 \times 10^2$  &  2  &   1152(1)  &  576(1)  &  40000  & 20000 &  1.352  & 0.150   & 0.7 \\  
  0.36 &$4 \times 10^8$ &$5.8 \times 10^2$ &  2  &   1152(1)  &  576(1)  &  20000  & 10000 &  6.003 & 0.642   & 0.2 \\  
  0.36 & $1 \times 10^9$ &$1.4 \times 10^3$ &  2 &   1152(3)  &  576(3)  &  12000  & 10000 &  15.10  & 1.632   & 0.5  \\  
  0.36 & $1.5 \times 10^9$ &$2.2 \times 10^3$ &  2 &   1152(3)  &  576(3)  &  10000  & 10000 &  22.15  & 2.418   & 0.8  \\  
  0.36 &$2 \times 10^9$ &$2.9 \times 10^3$ &  2  &   1152(3)  &  576(3)   & 15000  & 5000 &  28.49  & 3.164  & 2.6  \\  
  0.36 &$4 \times 10^9$ &$5.8 \times 10^3$ &  2  &   1152(3)  &  576(3)   & 10000  & 5000 &  50.12  &  5.958  & 3.3 \\ 
  0.36 & $1 \times 10^{10}$ &$1.4 \times 10^4$ &  2 &   1280(4)  &   640(4)  & 10000  &  3000 &  88.69  & 13.04   & 0.9  \\  
  0.36 &$2 \times 10^{10}$ &$2.9 \times 10^4$ &  2  &   1280(4)  &   640(4)   &  5000  &  2000 &  124.7  & 21.90  & 1.5  \\  
  0.36 &$4 \times 10^{10}$ &$5.8 \times 10^4$ &  2  &   1280(4)  &   640(4)   &  5000  &  2000 &  164.9  &  35.62  & 1.0  \\  
  0.36 & $1 \times 10^{11}$ &$1.4 \times 10^5$ & 2  &   1536(4)  &   768(4)   &  5000  &  2000  &  218.5  &  59.78  & 0.7  \\
  0.36 & $4 \times 10^{11}$ &$5.8 \times 10^5$ & 2  &   1536(4)  &   768(4)   &  4000  &  2000  &  374.1  &  158.5  & 0.5  \\
  0.36 & $1 \times 10^{12}$ &$1.4 \times 10^6$ & 1  &   1536(4)   &  1536(4)  &  1000  &  500  &  488.0  &  279.8  & 0.1  \\  [5pt]

  0.15 &$1 \times 10^{10}$ &$6.4 \times 10^2$ & 1 &   1536(1)   &  1536(1)  &  10000  &  4000 &  9.427  &  2.201  & 0.1 \\ 			
  0.15 &$2 \times 10^{10}$ &$1.3 \times 10^3$ & 1 &   1536(1)   &  1536(1)  &  14000  &  6000 &  17.23  &  4.269 & 0.1   \\ 
  0.15 &$3 \times 10^{10}$ &$1.9 \times 10^3$ & 1 &   1536(1)   &  1536(1)  &  14000  &  8000 &  25.25  &  6.297 & 0.3   \\ 
  0.15 &$4 \times 10^{10}$ &$2.6 \times 10^3$ & 1 &   1536(1)   &  1536(1)  &  10000   &  8000 &  33.74  &  8.480  & 0.4  \\ 			
  0.15 &$1 \times 10^{11}$ &$6.4 \times 10^3$ & 1 &   1536(2)   &  1536(2)  &  5000  &  2000 &  77.71  &  20.41  & 0.8   \\ 
  0.15 &$4 \times 10^{11}$ &$2.6 \times 10^4$ & 1 &   1536(4)   &  1536(4)  &  2000  &  2000 &  190.3  &  61.18  & 2.4  \\ 
  0.15 &$1 \times 10^{12}$ &$6.4 \times 10^4$ & 1 &   1536(4)   &  1536(4)  &  2000  &  3000 &  311.6  &  123.0  & 0.2  \\ 
  0.15 &$2 \times 10^{12}$ &$1.3 \times 10^5$ & 1 &   2048(4)   &  2048(4)  &  1500  & 1000 &  388.0  & 194.2 & 1.9  \\ 
  0.15 &$4 \times 10^{12}$ &$2.6 \times 10^5$ & 1 &   2048(4)   &  2048(4)  &  1000  &  1000 &  566.9  &  324.9  & 0.4  \\
\end{tabular}
\caption{Numerical details for all the cases. Columns from left to right are: the porosity, the Rayleigh number, the Rayleigh-Darcy number, the aspect ratio, the resolutions with refined factors in the horizontal and vertical directions, the simulation time before the statistical stage, the statistical time, the statistical Sherwood number, the statistical Reynolds number and the relative difference of the statistical Sherwood numbers calculated at the two plates.  }
\label{tab:cases}
\end{center}
\end{table}
We conduct direct numerical simulations (DNS) by solving the governing equations \eqref{eq:ns}, together with the additional body force $\boldsymbol{f}$ mentioned above. The code used in this study is an improved version based on our in-house code, which has been widely used in research on double-diffusive convection \citep{yang2015,yang2022,li2023wall}. Basically, the finite difference method and the fractional time-step method are employed \citep{ostilla2015}. One advantage of this code is the use of a dual-resolution technique, where a more refined grid is employed for the scalar field with high Schmidt number $\Sch$. In our previous studies, $\Sch$ reached as high as 1000 \citep{yang2022}. In the current study, we fix $\Sch=400$, a value close to the practical conditions of CO$_2$ sequestration \citep{huppert2014fluid}. For the three types of porous medium configurations shown in table \ref{tab:porous}, we have set a wide range of Rayleigh number $\Ray$, with the corresponding Rayleigh-Darcy number $\Ray_D$ ranging from $10^2$ to $10^6$. In the classical Darcy model, $\Ray_D >10^4$ signifies the ultimate regime \citep{nield2017}. Therefore, our parameter range extends beyond this regime to study the transition from the Darcy regime to the non-Darcy regime. The details of the simulations are summarized in table \ref{tab:cases}.

For most cases with $\phi=0.64$ and $0.36$, we set the aspect ratio $\Gamma=2$; for cases with $\phi=0.36$ and $\Ray=10^{12}$, as well as all cases with  $\phi=0.15$, we set $\Gamma=1$ to save computational resources. The grid resolution meets the following three criteria: First, the gap between two obstacles must be at least five grid cells, implying the base grid scale is less than $0.2l$; second, the base grid scale is smaller than the Kolmogorov scale $\eta_K=(\nu^3/\epsilon)^{1/4}$ \citep{grotzbach1983spatial}, in which $\epsilon$ is the mean viscous dissipation rate; third, the refined grid scale is smaller than the Batchelor scale $\eta_B=(\nu \kappa^2/\epsilon)^{1/4}$ \citep{silano2010numerical}. For all cases, we simulated for a sufficiently long time $t_d$ to ensure that the flow reached a statistically steady state. Then, we continued the simulation for an period $t_s$ for statistical analysis. The relative difference between the time-averaged Sherwood number $\overline{\She}$  in the first and second halves of this period did not exceed $1\% $. Hereafter, the overline denotes time-averaged values. Additionally, we calculated the relative difference of $\overline{\She}$ on the upper and lower plates, which was less than $4\%$ for all cases. This indicates that no additional mass flux entered or exited the obstacles.


\section{Flow structure}\label{sec:structure}
\subsection{Transition of the flow field}
\begin{figure}
 \centering
 \includegraphics[width=1\textwidth]{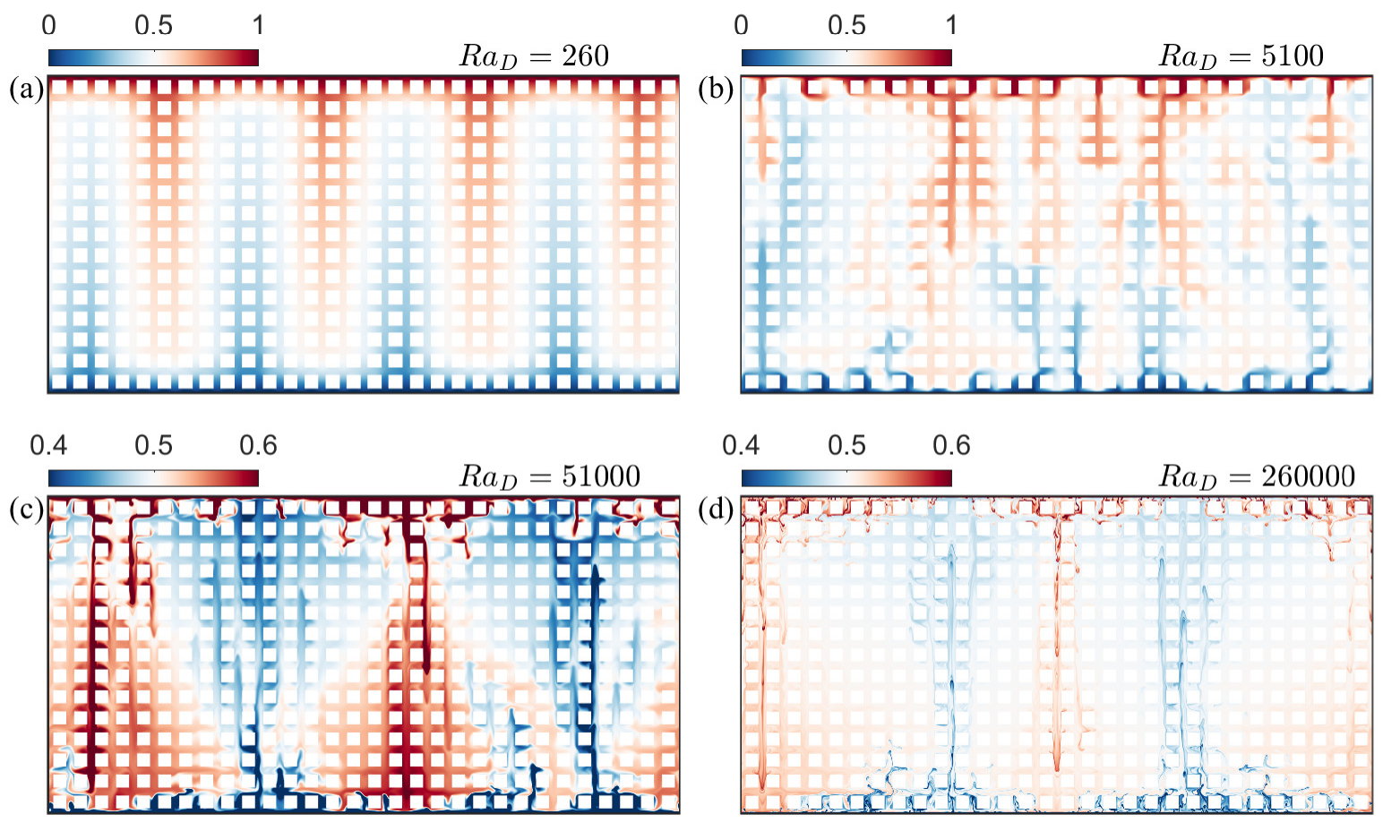}%
 \caption{Snapshots of the instantaneous concentration fields for the cases with (a) $\Ray=10^7$, (b) $\Ray=2 \times 10^8$, (c) $\Ray=2 \times 10^9$ and (d) $\Ray=10^{11}$. The porosity is fixed at 0.64. The obstacles are denoted by the white squares.}
 \label{fig:flowfield}
\end{figure}
In this section, we examine the characteristic structures of the flow field under varying $\Ray$ or $\Ray_D$ . Based on the Darcy model, convection in porous media can be classified into five regimes \citep{nield2017,Gasow2020}: the conducting regime ($0 \leq \Ray_D \leq 4\pi^2$), the steady state regime ($4\pi^2 \leq \Ray_D \leq 350$), the quasi-periodic regime ($350 \leq \Ray_D \leq 1300$), the high Rayleigh regime ($1300 \leq \Ray_D \leq 10000$) and the ultimate Rayleigh regime ($\Ray_D \geq 10000$). Figure \ref{fig:flowfield} presents four typical concentration fields for the cases of $\phi=0.64$. When $\Ray=10^7$ and $\Ray_D=260$, the flow field is composed of four pairs of convection cells, corresponding to the steady state regime, as shown in figure \ref{fig:flowfield}a. At this stage, the concentration distribution in the region near the upper and lower boundary plates is very uniform over at least one REV. As the Rayleigh number increases, the large-scale structures in the flow field begin to become disordered, entering the high Rayleigh regime, as shown in figure \ref{fig:flowfield}b. At this stage, the large-scale convection cells disappear, replaced by smaller-scale plumes that are chaotically distributed. These plumes originate from the upper and lower boundaries and merge into larger-scale plumes in the middle of the domain. When $\Ray_D$ further increases over 10000, the flow enters the ultimate Rayleigh regime. As shown in figure \ref{fig:flowfield}c, the characteristic structures of the concentration field comprise two parts: one consists of two pairs of large-scale convection cells with a fan-like shape probably caused by mechanical dispersion \citep{wen2018rayleigh}, and the other consists of small-scale plumes generated at the boundaries, with sizes smaller than the gaps between obstacles. Finally, for the highest Rayleigh number in this set of cases, i.e. $\Ray=10^{11}$ and $\Ray_D=260000$, the bulk area is thoroughly mixed, with clearer convection cells, as shown in figure \ref{fig:flowfield}d. Meanwhile, the extremely small-scale plumes near the boundaries become highly turbulent. The flow field at this stage resembles classical RB convection, with obstacles having minimal impact.

\begin{figure}
 \centering
 \includegraphics[width=1\textwidth]{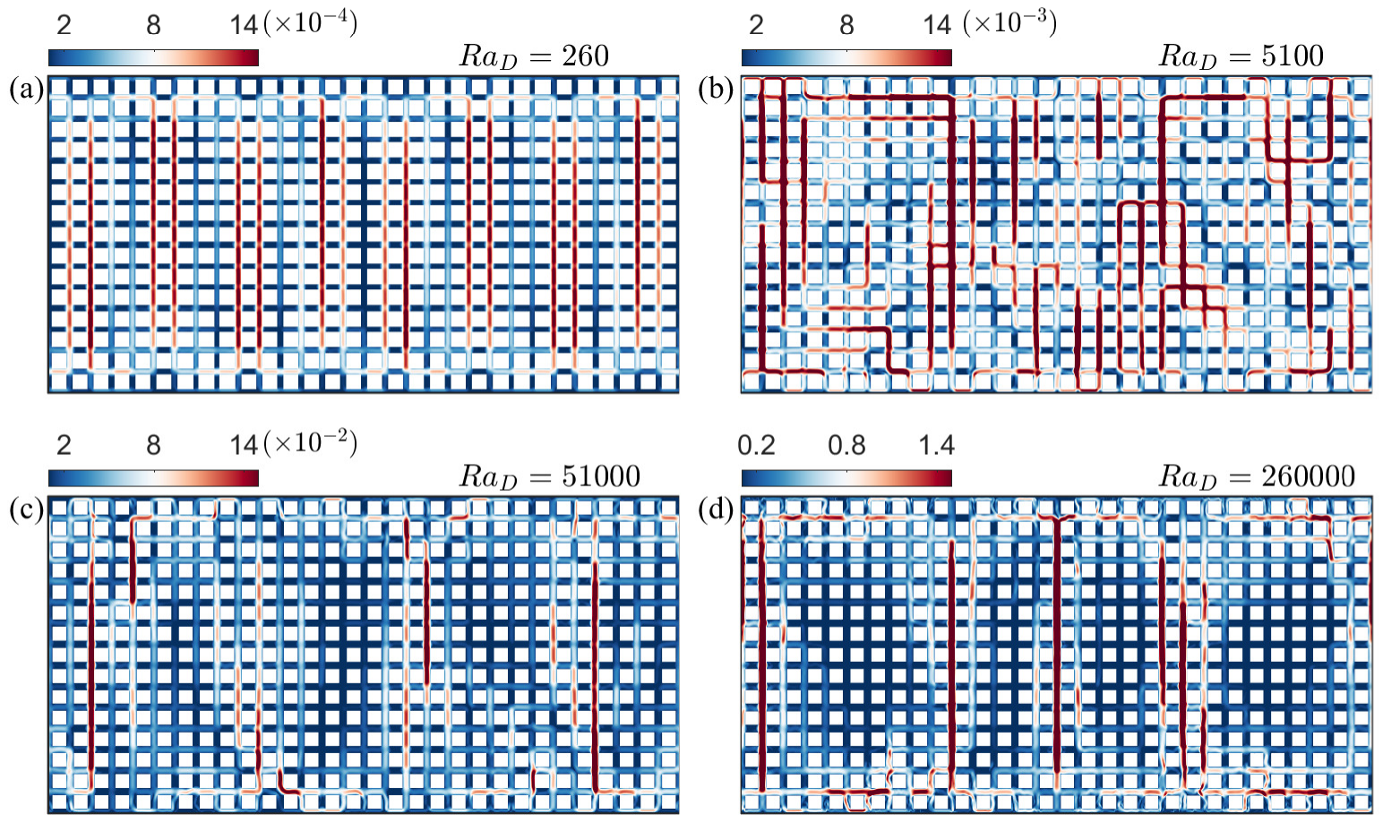}%
 \caption{Snapshots of the instantaneous pore-scale Reynolds number $\Rey_K$ for the cases with (a) $\Ray=10^7$, (b) $\Ray=2 \times 10^8$, (c) $\Ray=2 \times 10^9$ and (d) $\Ray=10^{11}$. The porosity is fixed at 0.64. The obstacles are denoted by the white squares.}
 \label{fig:flowfield-Rek}
\end{figure}
It is important to note that the chaotic structures in figure \ref{fig:flowfield}b differ from classical turbulence, with a very low Reynolds number of about 2 (see table \ref{tab:cases}). In the field of porous media convection, this type of flow is often referred to as ``pseudo-turbulence" \citep{jin2024}. By defining the pore-scale Reynolds number, i.e.
\begin{equation}
 \Rey_K = \frac{|\boldsymbol{u}^*| \sqrt{K^*}}{\nu}=\sqrt{\frac{\Ray_D}{\Sch}}|\boldsymbol{u}|,
\end{equation}
we can better characterize the turbulent flows in porous media. Here, $\sqrt{K^*}$ is proportional to the size $D^*$ of REV according to the equation \eqref{eq:K}. Therefore, $\Rey_K$ represents the ratio of inertial to viscous dissipation at the pore scale. A Darcy-type flow occurs when $\Rey_K\ll 1$\citep{nield2017,Gasow2020}. Figure \ref{fig:flowfield-Rek} presents the contour plots of $\Rey_K$ corresponding to four instantaneous fields shown in figure \ref{fig:flowfield}. It can be observed that in the first two cases, the magnitude of $\Rey_K$ is much less than 1, indicating that the flow in the pores is not turbulent and conforms to Darcy flow. However, for the flow field in figure \ref{fig:flowfield-Rek}b, the larger-scale structures are random and chaotic, leading us to classify this as ``pseudo-turbulence''. For the cases in figures \ref{fig:flowfield-Rek}c and \ref{fig:flowfield-Rek}d, the maximum of $\Rey_K$ approach or exceed 1, clearly indicating that the classical Darcy model is no longer applicable at this stage. Furthermore, we can examine the transition process from Darcy flow to non-Darcy flow through the distribution of $\Rey_K$. At $\Ray_D=260$, the flow field consists of relatively stable convection cells, with alternating regions of higher and lower velocities in the bulk. When $\Ray_D=5100$, the chaotic distribution of $\Rey_K$ clearly represents the pseudo-turbulent state at this stage. At $\Ray_D=51000$, two pairs of large-scale convection cells form, with higher velocities primarily appearing at the boundaries of the cells, while the velocities in their centers are very low, as seen in the dark areas of figure \ref{fig:flowfield-Rek}c. At this stage, some smaller-scale plumes still appear in the bulk region. For $\Ray_D=260000$, the bulk region is essentially dominated by only two pairs of convection cells, with very low velocities in the large central areas of these cells.

\subsection{Horizontal characteristic scale}
We have observed that as the Rayleigh number increases, the flow state gradually transitions from Darcy flow to non-Darcy flow, with the transitional state exhibiting complex flow structures. In addition to the pore-scale Reynolds number being sufficiently small, the condition for the existence of Darcy flow also requires that the characteristic scale of the flow field structure be much larger than the size of REV \citep{hewitt2020vigorous,de2023convective}. It is noteworthy that the horizontal scale of plumes is generally much smaller than their vertical scale. Therefore, in this section, we further examine the variations in the horizontal characteristic scale of the flow field.

\begin{figure}
 \centering
 \includegraphics[width=1\textwidth]{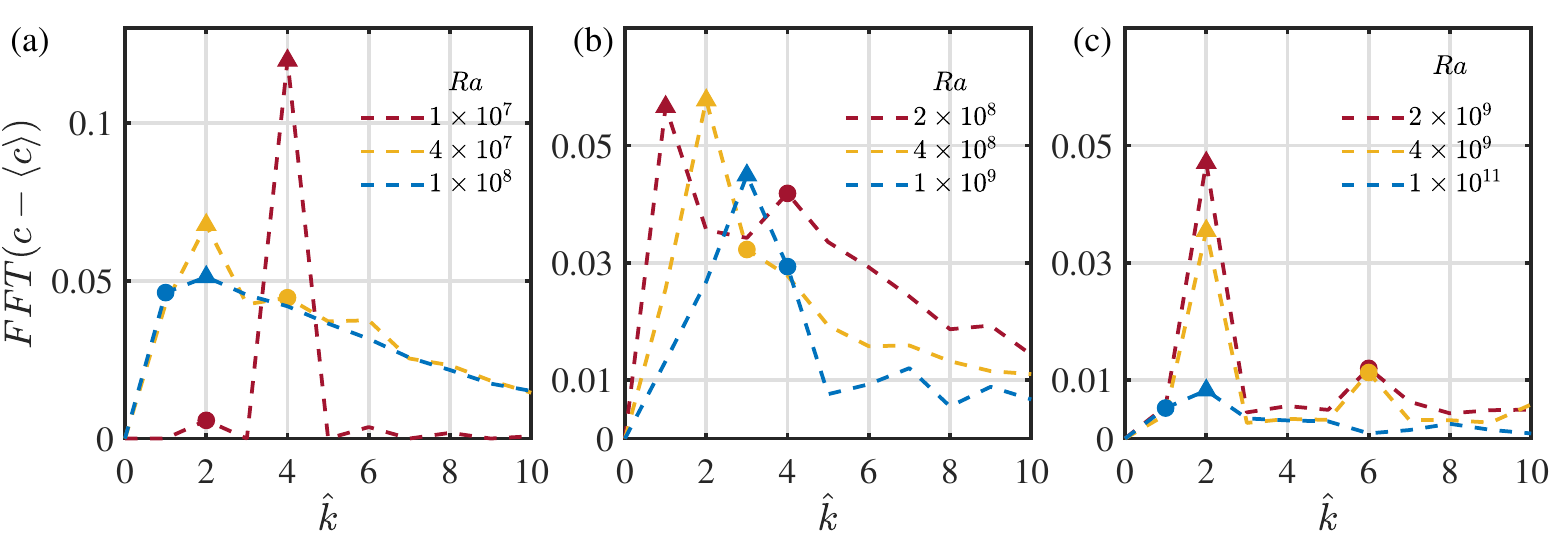}%
 \caption{Time-averaged spectra of the mass concentration at $ d/2 \leq|z-0.5| \leq 2/D$. The sampled data for Fast Fourier Transform (FFT) comes from the statistical steady state of the cases with $\phi=0.64$. The triangles indicate the first dominant wavenumber, while the circles indicate the second dominant wavenumber. The wavenumber is re-scaled as $\hat{k}=k_x\Gamma/2\pi$.}
 \label{fig:FFT}
\end{figure}
First, a simple method to find the characteristic wavelength $\lambda_x$, or the corresponding wavenumber $k_x$, is the Fast Fourier Transform (FFT). Figure \ref{fig:FFT} shows the horizontal wavenumber spectra for all cases with $\phi=0.64$. During sampling, we selected data from the middle height region of the domain to avoid the influence of the upper and lower boundaries. Since obstacles are arranged at $z=0.5$, we selected the interval of $ d/2 \leq|z-0.5| \leq 2/D$, which is a region within the height of one REV, excluding the obstacles. Additionally, for each case, we performed FFT on the instantaneous concentration field for at least 100 time points and then averaged them over time. The wavenumber is re-scaled as $\hat{k}=k_x\Gamma/2\pi$ thus it directly represents the number of periodic structures within the domain width. In figure \ref{fig:FFT} we use triangles and circles to denote the first and the second dominant wavenumbers, respectively. From figure \ref{fig:FFT}a, we can see that for the smallest Rayleigh number ($\Ray=10^7$) considered here, the wavenumber $\hat{k}=4$ is absolutely dominant, corresponding to the stable four pairs of convection cells in figure \ref{fig:flowfield}a. However, when $\Ray$ increases to $4\times10^7$ and $10^8$, $\hat{k}=2$ replaces $\hat{k}=4$ as the dominant wavenumber, with their proportions being relatively close. Additionally, other small wavenumbers, such as $\hat{k}=1$ and $\hat{k}=3$, also have significant shares. This indicates that the flow field is tending towards a more chaotic state. When $\Ray$ increases to $2\times10^8$, the first dominant wavenumber is $\hat{k}=1$, corresponding to the highly chaotic and random flow state in figure \ref{fig:flowfield}b, where there is no obvious periodic structure. For the three cases in figure \ref{fig:FFT}b, the shares of the small wavenumbers are relatively even, indicating that there is no stable periodic structure in the flow field at this stage, and plumes of multiple scales are distributed chaotically in both time and space. When $\Ray$ further increases to $2\times 10^9$, the wavenumber $\hat{k}=2$ becomes dominant, and its proportion far exceeds that of other wavenumbers, as shown in figure \ref{fig:FFT}c. This corresponds to the two pairs of convection cells in figure \ref{fig:flowfield}c. After this, as $\Ray$ increases, the dominant wavenumber keeps at $\hat{k}=2$, and the intensity of all the modes gradually decreases. When $\Ray$ reaches $10^{11}$, the intensity of the dominant wavenumber is quite low, implying the weak disturbance in the concentration field, as shown in figure \ref{fig:flowfield}d. Overall, as $\Ray$ or $\Ray_D$ increases, the dominant wavenumber of the flow field exhibits a relatively complex, non-monotonic change, with multiple wavenumbers interacting with each other. Previous numerical simulations under the Darcy model indicate $k_x\sim \Ray_D^{0.4}$ \citep{hewitt2014high}. In the current pore-scale simulations, due to the interaction between large-scale convection cells and small-scale plumes, FFT alone cannot provide similar conclusions.

\begin{figure}
 \centering
 \includegraphics[width=1\textwidth]{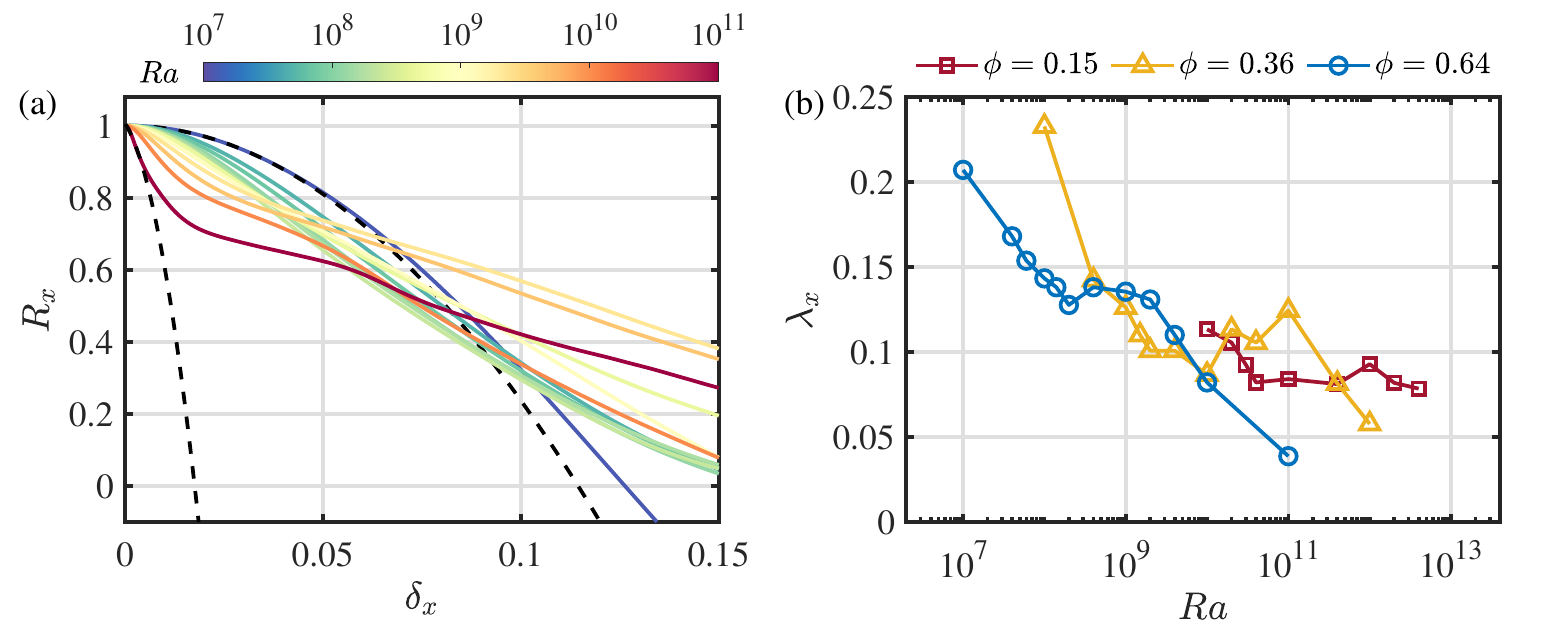}%
 \caption{(a) The horizontal auto-correlation function $R_x(\delta_x)$ of the mass concentration at $ d/2 \leq|z-0.5| \leq 2/D$. The sampled data comes from the statistical steady state of each case with $\phi=0.64$. The dashed lines denote the quadratic fitting of $R_x$ within $R_x\geq0.9$. (b) The horizontal characteristic wavelength $\lambda_x$ versus the Rayleigh number $\Ray$. $\lambda_x$ is four times the $\delta_x$ value at $R_x=0$ from the quadratic fitting curve shown in panel a.}
 \label{fig:calscale}
\end{figure}
To uniformly examine the variation pattern of the characteristic scale, we calculated the horizontal auto-correlation coefficient of the concentration field within the middle height region, namely:
\begin{equation}
 R_x(\delta_x)=\frac{\langle \overline{c(x,z,t)c(x+\delta_x,z,t)} \rangle_{z_{mid}}}{\langle\overline{ c^2(x,z,t)}\rangle_{z_{mid}}}.
\end{equation}
For $\phi=0.64$ and $0.15$, $z_{mid}$ is $d/2 \leq|z-0.5| \leq D/2$; for $\phi=0.36$, $z_{mid}$ is $|z-0.5| \leq l/2$. Figure \ref{fig:calscale}a plots $R_x(\delta_x)$ for all cases with $\phi=0.64$. The colors denote different $\Ray$. It can be seen that when $\Ray=10^7$, $R_x$ rapidly decreases from 1 to below 0 as $\delta$ increases, without any inflection points. This corresponds to the periodic structure shown in figure \ref{fig:flowfield}a. If we assume it satisfies a sinusoidal model \citep{hewitt2014high}, i.e., $c=c_0(z,t)sin(2\pi x/\lambda_x)$, then one can easily find $R_x(\lambda_x/4)=0$. This means we can extract four times the $\delta_x$ value at $R_x=0$ as the horizontal wavelength. However, as $\Ray$ increases, $R_x$ gradually deflects above 0. When $\Ray$ reaches around $10^9$, a clear inflection point appears at around $R_x=0.8$, corresponding to the emergence of large-scale structures (see figure \ref{fig:flowfield}c). Before the inflection point, $R_x$ still reflects the characteristics of small-scale structures to some extent. Therefore, we extract the $R_x$ values between 0.9 and 1 and perform a quadratic fitting, as shown by the dashed lines in figure \ref{fig:calscale}a. The $\delta_x$ value at $R_x=0$ from the fitted curve can be approximately considered as a quarter of the small-scale structure wavelength. Using this method, we calculated the horizontal wavelengths for all cases, with the results plotted in figure \ref{fig:calscale}b. For three sets of cases with different porosities, the variation of $\lambda_x$ with $\Ray$ is similar. That is, as $\Ray$ increases, $\lambda_x$ first decreases rapidly, then remains unchanged, and even slightly increases, before continuing to decrease. The intermediate plateau phase reflects the complex transition of the flow regime from Darcy flow to RB convection. The increase in scale with increasing $\Ray$ might be due to dispersion effects, which is also found in experiments \citep{liang2018effect}.

\begin{figure}
 \centering
 \includegraphics[width=1\textwidth]{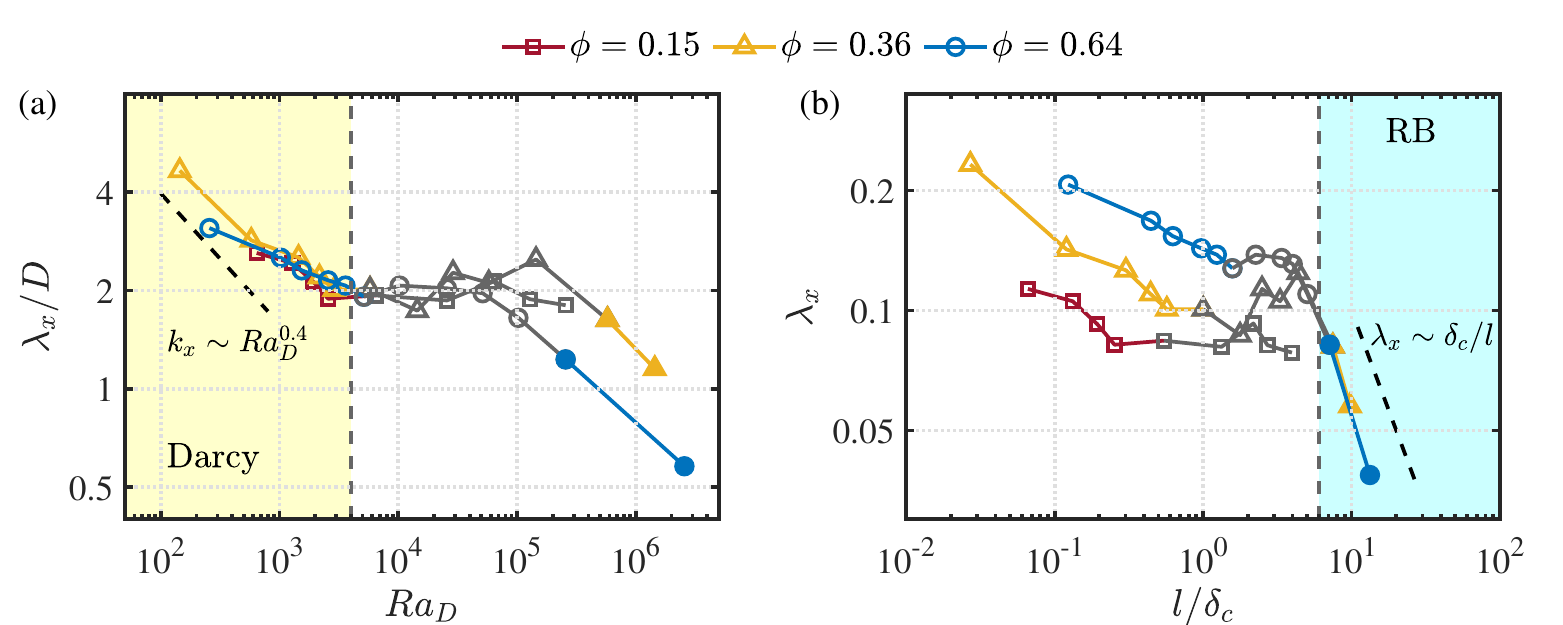}%
 \caption{(a) The horizontal characteristic wavelength $\lambda_x$ re-scaled by the REV scale $D$ versus the Rayleigh-Darcy number $\Ray_D$. The vertical dashed line denotes $\Ray_D=4000$. (b) $\lambda_x$ versus the ratio of the pore space $l$ and concentration BL thickness. The vertical dashed line denotes $l/\delta_c=6$.  In both panels, the open coloured symbols denote the Darcy regime, the grey symbols denote the transitional regime, and the solid symbols denote the RB regime. }
 \label{fig:scale}
\end{figure}
Previous pore-scale simulations \citep{Gasow2020} indicate that the scale of interior plumes in the Darcy regime increases with increasing $D$, namely the scale of REV. In figure \ref{fig:scale}a, we show the variation of $\lambda_x/D$ with $\Ray_D$. It can be seen that when $\Ray_D$ is small, the cases with different $\phi$ largely overlap; and as $\phi$ decreases, the wavenumber $k_x=2\pi/\lambda_x$ and $\Ray_D$ gradually approach the $0.4$ scaling \citep{hewitt2014high,Gasow2020}, as indicated by the black dashed line in the figure. When $\Ray_D$ reaches around 4000, $\lambda_x/D$ starts to stabilize at around 2 and does not continue to decrease, implyting that the width of a single plume ($\lambda_x/2$) remains comparable to the scale of one REV. This indicates that when the characteristic scale of the flow field structure decreases to the REV scale, the Darcy flow assumption starts to fail. Over a considerable range of $\Ray_D$ values thereafter (up to around $\Ray_D=10^5$), although the intrinsic scale of the plumes decreases, the dispersion effect caused by the porous structure leads to its actual scale always increasing to the REV scale. We define the flow state at this stage as being in the transition regime, indicated by grey symbols in figure \ref{fig:scale}. Before this stage, the variation in characteristic scale follows the features of Darcy flow, so we refer to the flow state at this stage as being in the Darcy regime.

After the transition regime, $\lambda_x$ begins to decrease continuously with the increase of $\Ray_D$. At this stage, the cases with different $\phi$ separate, indicating that the characteristic scale is no longer controlled by $D$. Similarly, $\Ray_D$, as the control parameter describing the Darcy regime, is no longer applicable at this stage. From figure \ref{fig:flowfield}, we can observe that the flow field enters a state similar to RB convection. Our previous work found that when the thermal boundary layer (BL) thickness is less than the pore space $l$, the heat transport begins to follow the classic $0.3$ scaling in RB convection \citep{liu2020rayleigh}. Therefore, we also examine the relationship between the concentration BL thickness $\delta_c$ and $l$ in the current cases. Here, the dimensionless thickness is defined as $\delta_c=\delta_c^*/H^*=1/(2\She)$. Figure \ref{fig:scale}b shows the relationship between $\lambda_x$ and $l/\delta_c$. We find that the transition regime does not end at $l/\delta_c=1$, but around $l/\delta_c=6$, as indicated by the grey dashed line in the figure. After this, the cases with different $\phi$ overlap again and satisfy the linear relationship $\lambda_x \sim \delta_c/l$. We refer to the flow state that satisfies this relationship as being in the RB regime, indicated by solid symbols. In the next section, we will show that $\She$ and $\Ray$ indeed satisfy the 0.3 scaling of RB convection in this regime.

\begin{figure}
 \centering
 \includegraphics[width=1\textwidth]{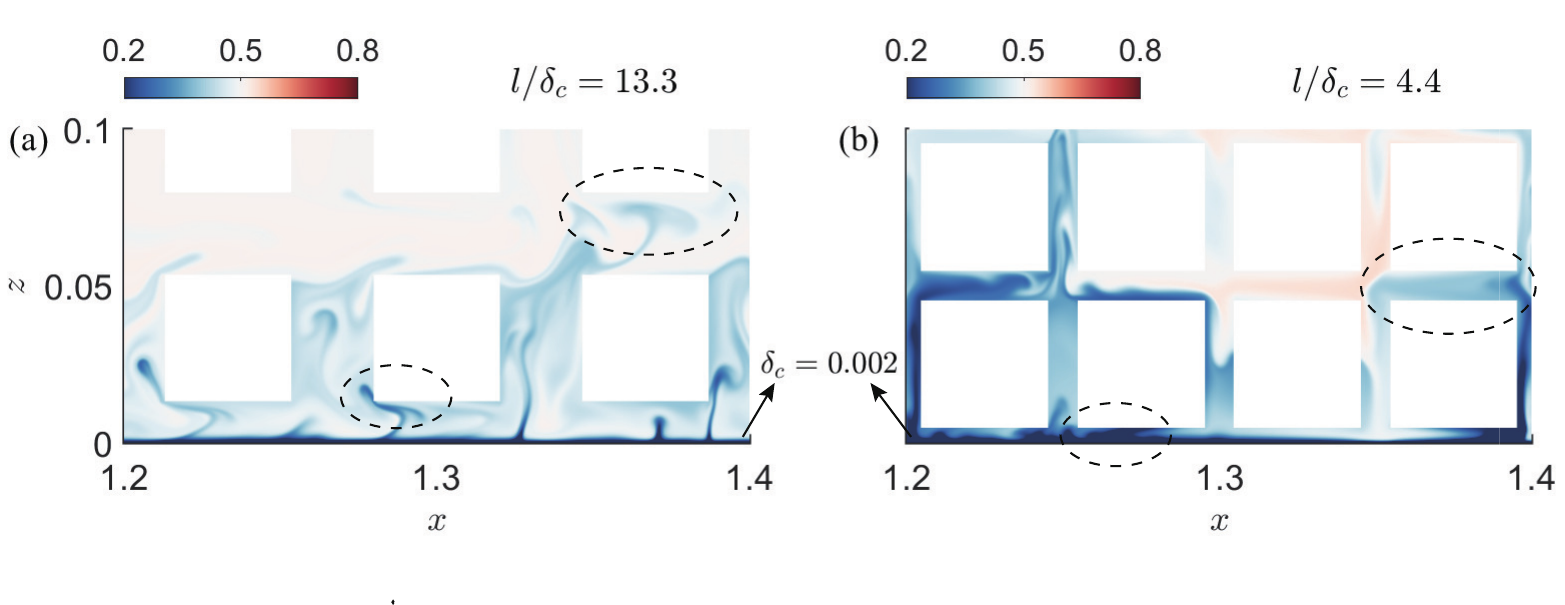}%
 \caption{The zoom-in plots of the instantaneous concentration fields near the bottom plate. The two cases have same $\Ray=10^{11}$ and $\delta_c=0.002$, but with different porosities: (a) $\phi=0.64$, (b) $\phi=0.36$. The plumes within the dashed ellipse undergo mechanical dispersion due to the obstacles. }
 \label{fig:flowfield-plume}
\end{figure}
It is natural that $\lambda_x$ and $\delta_c$ satisfy a linear relationship because the scale of the plume shedding from the boundary layer is similar to the BL thickness \citep{van2015plume}. But why is $\lambda_x$ inversely related to $l$? Figure \ref{fig:flowfield-plume} shows the instantaneous concentration fields near the bottom boundary for two cases with $\phi=0.64$ and $0.36$. These two cases have the same Rayleigh number $\Ray=10^{11}$, and their Sherwood numbers $\She$ are similar, resulting in almost the same boundary layer thickness $\delta_c$, both of which are smaller than the corresponding pore space $l$. From figure \ref{fig:flowfield-plume}a, we can see that when $l$ is much larger than $\delta_c$, the plumes can grow relatively freely. However, when they encounter obstacles, mechanical dispersion occurs, leading to an increase in the horizontal scale, as seen in the dashed ellipses in the figure. A naive idea is to assume that one plume maintain its volume when it encounters obstacles. If it has an intrinsic vertical scale $\lambda_z$, we have $\lambda_z \delta_c=\lambda_x l$, from which we can derive $\lambda_x \sim \delta_c/l$. When $l$ and $\delta_c$ are comparative, as shown in figure \ref{fig:flowfield-plume}b, the constraint on the plumes by the obstacles becomes very strong. At this stage, the plumes can disperse to the REV scale, and the flow field is still in the transitional regime. This explains why the transition to the RB regime requires $l$ to be much larger than $\delta_c$. Additionally, we should note that in the RB regime, although the porous structure no longer affects the BLs and therefore does not affect the transport efficiency, it still impacts the small-scale structure of the internal flow field.

\section{Mass transfer and energy dissipation}\label{sec:transfer}
\begin{figure}
 \centering
 \includegraphics[width=1\textwidth]{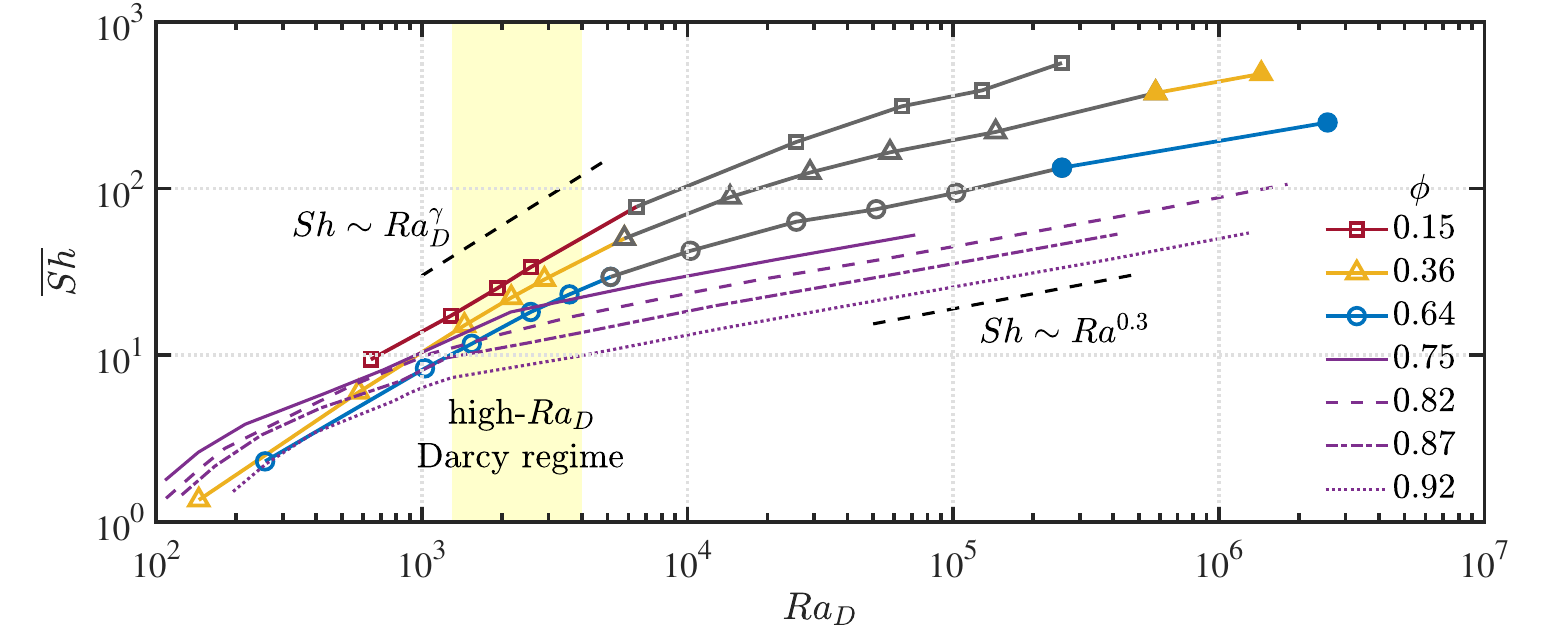}%
 \caption{The time-averaged Sherwood number $\overline{\She}$ versus the Rayleigh-Darcy number $\Ray_D$. The cases with $\phi=0.15-0.64$ are from the current study, where the obstacles are impermeable to the species and the Schmidt number is $\Sch=400$, while the cases with $\phi=0.75-0.92$ are from our previous study \citep{liu2020rayleigh}, where the obstacles can be penetrated by heat and the Prandtl number is $\Pra=4.3$. The open coloured symbols denote the Darcy regime, the grey symbols denote the transitional regime, and the solid symbols denote the RB regime. The yellow area denotes the high-$Ra_D$ Darcy regime with $1300 \leq Ra_D \leq 4000$; the exponent $\gamma$ is calculated by fitting the data within this area, which equals 0.81, 0.92 and 0.97 for $\phi=0.64$, 0.36 and 0.15, respectively.}
 \label{fig:regime}
\end{figure}
We are now examining the efficiency of mass transport in different regimes. In the Darcy model, classical theory posits that when flow enters the high Rayleigh regime (hereafter referred to as high-$\Ray_D$ Darcy regime), $\She$ and $\Ray_D$ satisfy a linear relationship, i.e., $\She \sim \Ray_D$ \citep{howard1966convection,doering1998bounds}. Recent two-dimensional Darcy-scale simulations have corroborated this conclusion well \citep{hewitt2014high}. However, both pore-scale simulations and experiments have observed nonlinear scaling. For example, \citet{Gasow2020}'s pore-scale simulations found $\She \sim \Ray_D^{0.8}$ for $\phi=0.56$, and \citet{liang2018effect}'s one-sided experiments found $\She \sim \Ray_D^{0.75}$. This indicates that the actual pore-scale structures can influence mass transfer. In the current study, we similarly observed nonlinear relationship for $\She$ and $\Ray_D$, as shown in figure \ref{fig:regime}. It can be seen that within the Darcy regime, the data approaches the linear scaling (indicated by the black dashed line). However, we performed a linear fit on data where $1300 \leq \Ray_D \leq 4000$, namely the high-$\Ray_D$ Darcy regime, and found $\She \sim \Ray_D^{0.81}$ for $\phi=0.64$, $\She \sim \Ray_D^{0.92}$ for $\phi=0.36$ and $\She \sim \Ray_D^{0.97}$ for $\phi=0.15$. These are consistent with the conclusions of \citet{Gasow2020,gasow2021,gasow2022}'s previous work. Nevertheless, with further increases in $\Ray_D$, our current cases reveal a transition from $\She \sim \Ray_D$ to $\She \sim \Ray^{0.3}$ (note that $\Ray_D \sim \Ray$ when $Da$ remains constant). 
This transition begins around $\Ray_D=4000$, even before the so-called ultimate Rayleigh regime. Additionally, within the transition regime, the scaling exponent of $\She$ with $\Ray_D$ gradually changes, making the initial phase of the transition regime potentially imperceptible. In the figure 8 of \citet{gasow2022}'s study, it can also be observed that as $\Ray_D$ approaches $10^5$, there is a trend of decreasing $\She$. The underlying physical interpretation for the transition may be the influence of porous structure on the flow structure. The transition starts when the characteristic scale of the flow structure is comparable to the size of REV. When the scale of the plumes, or the thickness of the boundary layer, is much smaller than the size of the pore space $l$, the effect of the porous media on mass transport can be neglected, and $\She$ with $\Ray$ satisfies the 0.3 scaling of RB convection. For smaller $\phi$ (corresponding to smaller $l$), the Rayleigh number required to reach the RB regime is larger. The current cases with $\phi = 0.15$ have not yet reached the RB regime, which would require $\Ray$ to exceed the order of $10^{13}$, posing a significant challenge under current computational conditions. Finally, we compare the results obtained from the previous work \citep{liu2020rayleigh} on thermal convection, as shown by the purple lines in the figure. Apart from the differences in Schmidt number, the obstacles in the earlier work are not adiabatic. This results in a much earlier transition, with the flow entering the RB regime at a relatively low Rayleigh-Darcy number.

\begin{figure}
 \centering
 \includegraphics[width=1\textwidth]{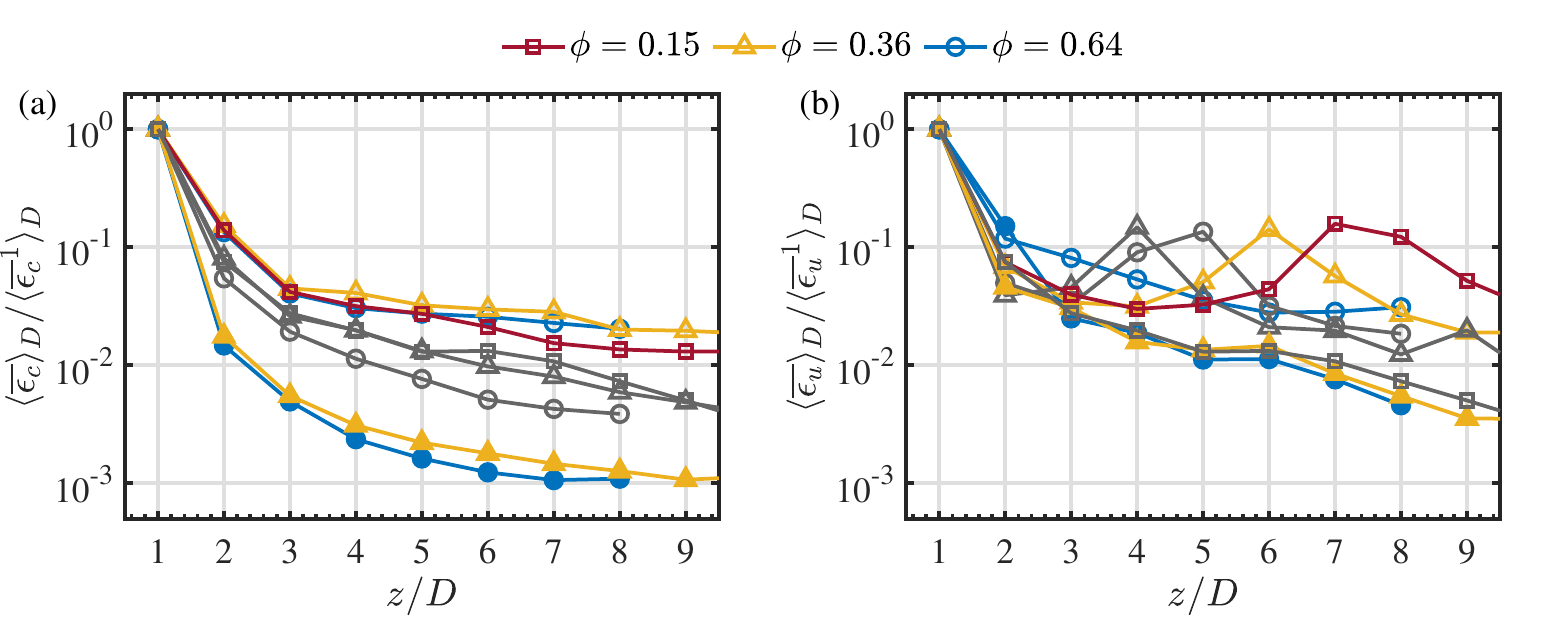}%
 \caption{Normalized dissipation rates for (a) concentration and (b) kinetic energy versus the distance from the wall. The distance is re-scaled by the size $D$ of REV. The dissipation rates are averaged both in time and space ($(n-1)D\leq z \leq nD, n=1,2,3...$). The open coloured symbols denote the cases with $Ra_D=2600$ ($\phi=0.64$ and 0.15) and $Ra_D=2900$ ($\phi=0.36$), the grey symbols denote the cases with $Ra_D=2.6 \times 10^4$ ($\phi=0.64$ and 0.15) and $Ra_D=2.9 \times 10^4$ ($\phi=0.36$), and the solid symbols denote the cases with $Ra_D=2.6 \times 10^6$ ($\phi=0.64$) and $Ra_D=1.4 \times 10^6$ ($\phi=0.36$).  }
 \label{fig:diss1}
\end{figure}
We further investigate the energy dissipation in different regimes. The dimensional energy dissipation rates for concentration and kinetic energy are defined as follows:
\begin{equation}
 \epsilon^*_c=\kappa\sum_i \left[ \frac{\partial c^*}{\partial x_i^*}\right]^2, \quad \epsilon_u^*=\frac{1}{2}\nu\sum_{ij} \left[ \frac{\partial u_j^*}{\partial x_i^*}+\frac{\partial u_i^*}{\partial x_j^*}\right]^2.
\end{equation}
The corresponding dimensionless forms read
\begin{equation}
 \epsilon_c=\frac{1}{\sqrt{RaSc}}\sum_i \left[ \frac{\partial c}{\partial x_i}\right]^2, \quad \epsilon_u=\frac{1}{2}\sqrt{\frac{Sc}{Ra}}\sum_{ij}
 \left[ \frac{\partial u_j}{\partial x_i}+\frac{\partial u_i}{\partial x_j}\right]^2.
\end{equation}
In figure \ref{fig:diss1}, we plot the vertical distribution of the two energy dissipation rates. Here, the energy dissipation rates are time-averaged and spatially averaged over vertical intervals of $D$, namely the size of REV. Additionally, the vertical coordinate is also re-scaled using $D$. For each regime and porosity, we selected cases with similar $\Ray_D$ values. From figure \ref{fig:diss1}a, it can be observed that, after normalizing by the value at the first point of each group (denoted by superscript 1), the distribution of $\epsilon_c$ for cases with similar $Ra_D$ is consistent. The dissipation rate rapidly decreases with increasing distance from the wall, and this decrease becomes more pronounced as $Ra_D$ increases. In the Darcy regime, $\epsilon_c$ decreases to approximately $4\%$ of $\epsilon_c^1$ at a distance of three REVs from the wall. For the transitional regime and RB regime, this ratio is approximately $2\%$ and $0.5\%$, respectively. Therefore, concentration energy dissipation is mainly concentrated within the distance of $3D$ from the wall. \citet{gasow2022} also reported that in their numerical simulations, energy dissipation within 5 REVs of the wall accounted for $93\%$ of the total dissipation. The distribution of kinetic energy dissipation, however, shows a different pattern. As shown in figure \ref{fig:diss1}b, at lower $\Ray_D$ values, there is a slight increase in $\epsilon_u$ near the center, reaching approximately $15\%$ of $\epsilon_u^1$. This is caused by the chaotic plumes observed in the high-$Ra_D$ Darcy regime and transitional regime. The numerous plume structures in the bulk area cause significant kinetic energy dissipation at the boundaries of obstacles. $\epsilon_c$ is less affected because the normal concentration gradient at the boundaries of obstacles is zero. Moreover, for different porosities, the change in $\epsilon_u$ with $Ra_D$ exhibits a lag effect. For larger porosities, as $Ra_D$ increases, $\epsilon_u$ rises later in the central region before eventually decreasing. In the RB regime, $\epsilon_u$ continues to decrease with increasing distance from the wall. At the fifth REV position, it has dropped to approximately $1\%$ of $\epsilon_u^1$.

\begin{figure}
 \centering
 \includegraphics[width=1\textwidth]{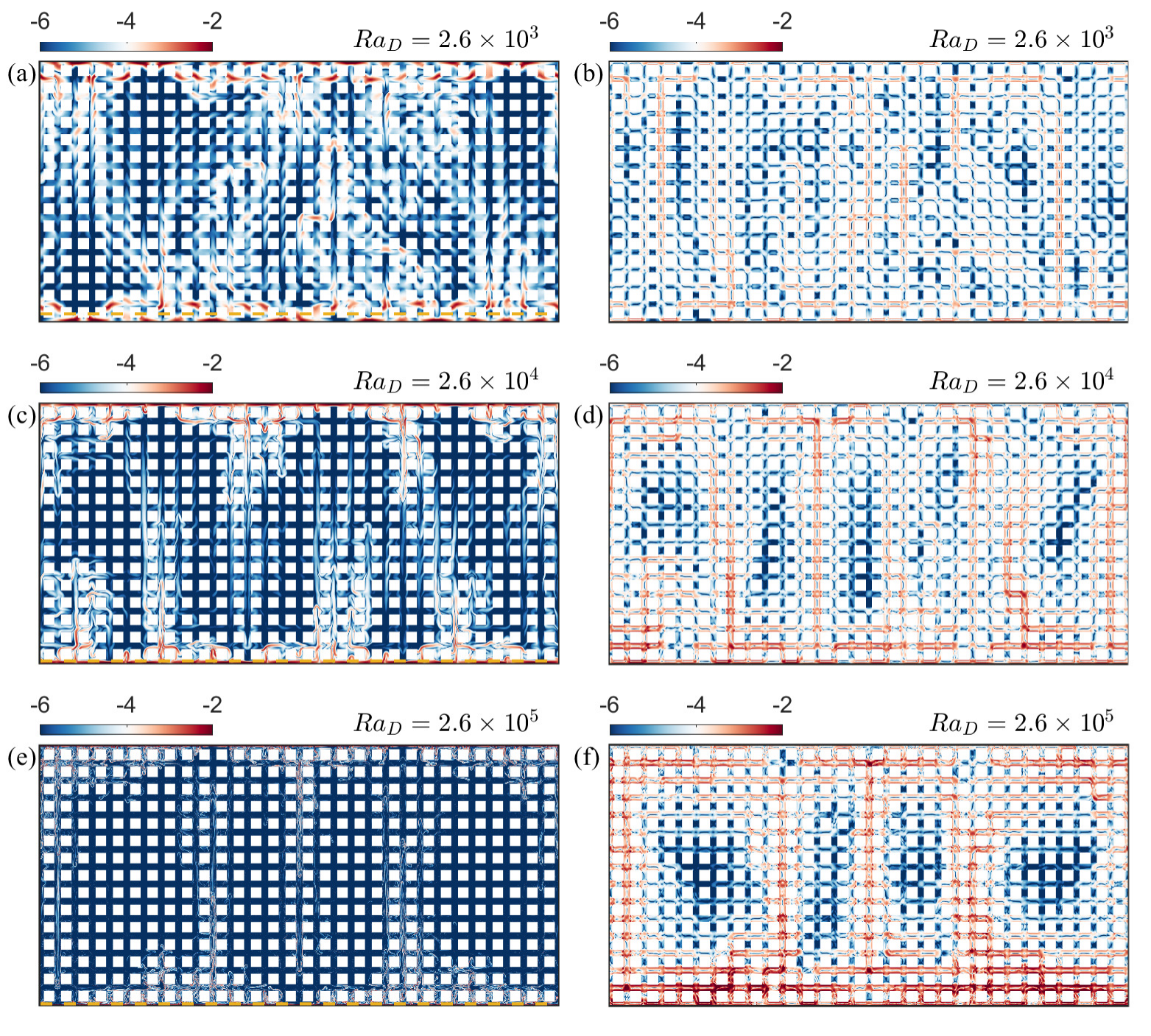}%
 \caption{Typical snapshots of (a, c, e) concentration energy dissipation rate log$_{10}\epsilon_c$ and (b, d, f) kinetic energy dissipation rate log$_{10}\epsilon_u$ for the cases with $\phi=0.64$. The yellow dashed lines in (a, c, e) denote the concentration BL. }
 \label{fig:diss2}
\end{figure}
The distribution of dissipation rates throughout the entire flow field allows us to see the differences between the various regimes more clearly, as shown in figure \ref{fig:diss2}. Here, we selected three cases represented by circular symbols ($\phi=0.64$) from figure \ref{fig:diss1}. In the high-$Ra_D$ Darcy regime (figures \ref{fig:diss2}a and \ref{fig:diss2}b), the chaotic plumes lead to an increase in both dissipation rates. For the concentration, dissipation is more pronounced near the boundaries and extends several REVs into the bulk region. In contrast, significant kinetic energy dissipation shows a larger region throughout the field, with most of the contribution coming from the plume structures interacting with the boundaries of obstacles. In the transitional regime (figures \ref{fig:diss2}c and \ref{fig:diss2}d), plume scales become smaller, and large-scale structures gradually form. This causes concentration energy dissipation to be more concentrated near the boundaries, with very low dissipation across a wide area in the bulk region. Conversely, for kinetic energy dissipation, the more intense flow in the bulk region leads to increased dissipation, even exceeding that near the boundaries. As the flow enters the RB regime (figures \ref{fig:diss2}e and \ref{fig:diss2}f), stable large-scale convective rolls result in an even larger low $\epsilon_c$ area within the bulk region, with dissipation mainly concentrated along the edges of the convective rolls. At this stage, kinetic energy dissipation throughout the field further increases, and contributions near the boundaries become more significant due to the presence of chaotic small-scale plumes near the boundaries.

In the next section, we will analyze the $Sh-Ra$ relationship based on Grossmann and Lohse (GL) theory \citep{grossmann2000scaling,grossmann2001thermal}, which primarily focuses on dividing energy dissipation rates into BL contributions and bulk contributions. Therefore, here we first examine the BL thickness. For the concentration field, we calculated the dimensionless BL thickness using $\delta_c=1/(2\She)$. In figures \ref{fig:diss2}(a, c, e), BLs are marked with yellow dashed lines. It can be observed that, in the Darcy regime, $\delta_c$ is approximately 0.028, exceeding the pore width $l$ but smaller than the REV size $D$. In this case, the dissipation rate within BLs is relatively large. In the transitional regime, $\delta_c$ is reduced to less than $l$, approximately 0.008. In the RB regime, $\delta_c$ becomes very small, around 0.004, resulting in a minimal contribution to the overall dissipation. For the kinetic BL, due to the no-slip condition at the obstacles, the BL thickness at both the upper and lower plates is always less than $l$. In our subsequent theoretical analysis, we consider that BLs exist at the boundaries of all obstacles throughout the field. The definitions of the bulk region and the BL region differ from the traditional ones in this context. Specific details will be introduced in the next section.

\section{Application of GL theory} \label{sec:GL}
The key idea of GL theory is to decompose the dissipation rates into contributions from the BLs and the bulk region, i.e.
\begin{equation}
 \epsilon_u^*=\epsilon_{u,BL}^*+\epsilon_{u,bulk}^*,\quad \epsilon_c^*=\epsilon_{c,BL}^*+\epsilon_{c,bulk}^*.
\end{equation}
Depending on the regime, the dissipation rates may be dominated by the BL part or the bulk part, leading to corresponding estimation formulas. Moreover, in a closed RB system, there exists exact relations \citep{grossmann2000scaling,grossmann2001thermal}:
\begin{equation}
 \langle \overline{\epsilon_u^*} \rangle_V=\frac{\nu^3}{H^{*4}}(\She-1)\Ray\Sch^{-2},\quad \langle \overline{\epsilon_c^*} \rangle_V=\kappa\frac{\Delta_c^*}{H^{*2}}\She. \label{eq:exact_relation}
\end{equation}
Here, the overline represents time-averaging, and the bracket $\langle \cdot \rangle_V$ indicates averaging over the entire domain. It should be noted that the presence of obstacles does not affect the validity of the exact relations \eqref{eq:exact_relation}. 

\begin{figure}
 \centering
 \includegraphics[width=1\textwidth]{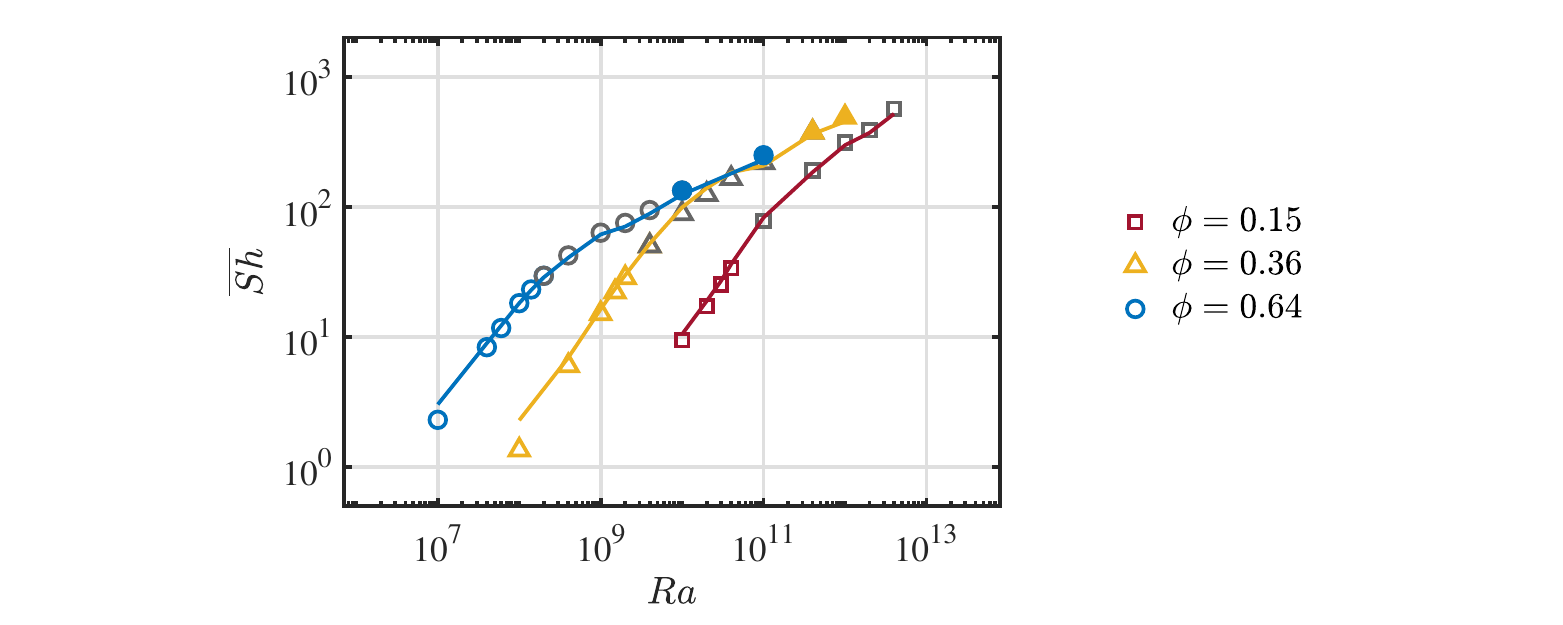}%
 \caption{The time-averaged Sherwood number $\overline{\She}$ versus the Rayleigh number $\Ray$. The open coloured symbols denote the Darcy regime, the grey symbols denote the transitional regime, and the solid symbols denote the RB regime. The solid lines denote the equation \eqref{eq:GL-Sh} with $\alpha=10$.}
 \label{fig:GL-Sh}
\end{figure}
In our previous work on thermal convection in porous media \citep{liu2020rayleigh}, we found that additional laminar-type BLs exist in the channels between obstacles, leading to the conclusion that kinetic energy dissipation is dominated by the numerous obstacle BLs throughout the domain. By combining the corresponding estimation formula with the exact relation \eqref{eq:exact_relation}, we obtain:
\begin{equation}
\She \approx \alpha \cdot \phi (\frac{H^*}{l^*})^2\Sch^2\Rey^2\Ray^{-1}+1. \label{eq:GL-Sh}
\end{equation}
Here $\alpha$ is an empirical coefficient. Note that we have replaced the Nusselt number $Nu$ and Prandtl number $Pr$ in the original formula (equation (5.3) in \citet{liu2020rayleigh}) with the Sherwood number $\She$ and Schmidt number $\Sch$, respectively, and the Reynolds number $\Rey$ is defined based on the domain height $H^*$ \eqref{eq:response}. More details can be found in that paper. Although this formula was derived for thermal convection, it theoretically also holds for the current cases with large $\Sch$. In figure \ref{fig:GL-Sh}, we plot the relationship between $\She$ and $\Ray$, along with equation \eqref{eq:GL-Sh} represented by the solid lines. Here, we set the coefficient $\alpha$ to 10 (compared to 8 in \citet{liu2020rayleigh}'s work). It can be observed that equation \eqref{eq:GL-Sh} matches the DNS results remarkably well across almost all regimes, including the high-$\Ray_D$ Darcy regime, transitional regime and RB regime. Together with previous work, equation \eqref{eq:GL-Sh} demonstrates high universality and is valid over a wide range of $\Sch$, $\Ray$, and $\phi$.

\begin{figure}
 \centering
 \includegraphics[width=1\textwidth]{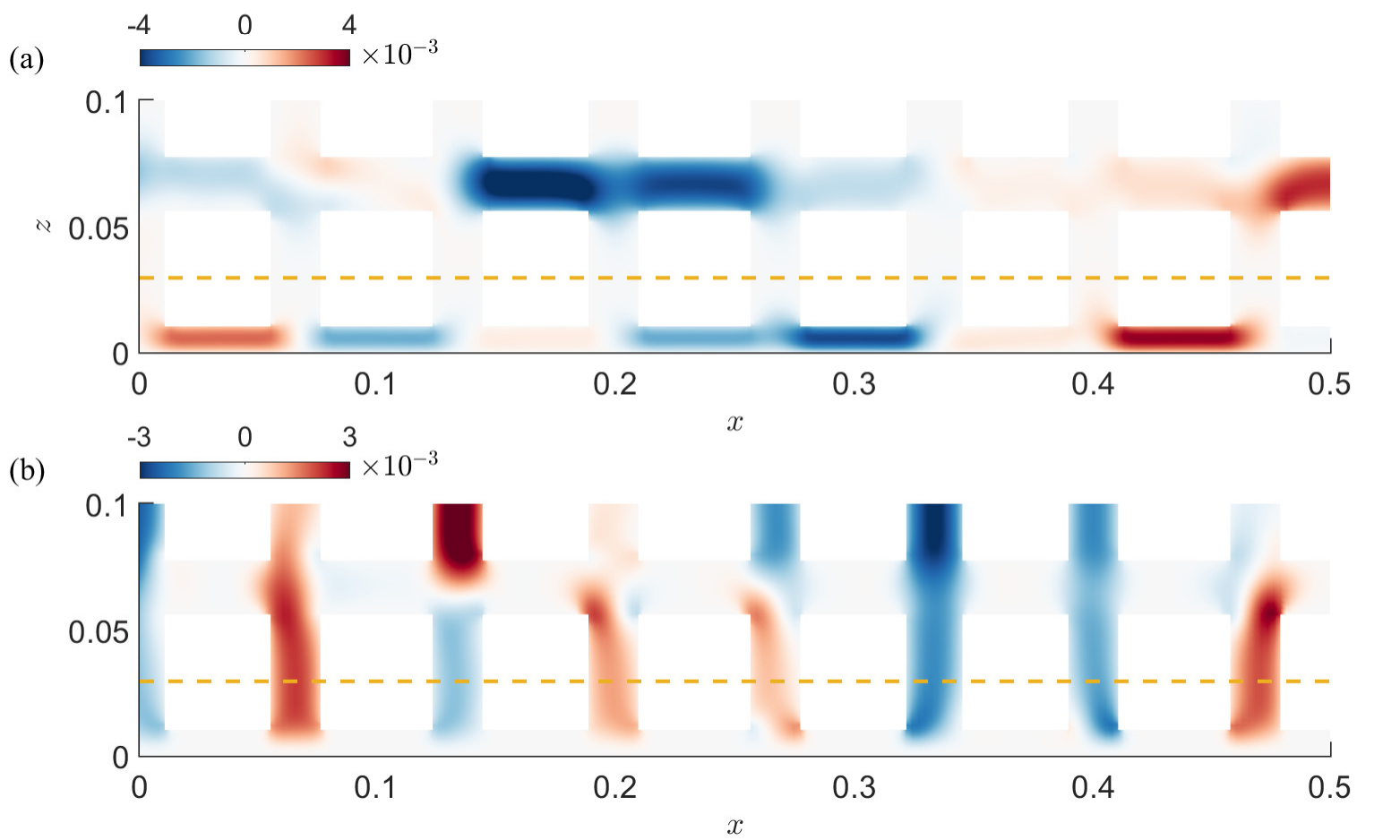}%
 \caption{The zoom-in plots of the instantaneous velocity fields near the bottom plate: (a) horizontal velocity $u_x$; (b) vertical velocity $u_z$. The case has $\Ray_D=2600$ and $\phi=0.64$. The dashed yellow lines denote the concentration BLs. }
 \label{fig:GL-BLvelocity}
\end{figure}

\begin{figure}
 \centering
 \includegraphics[width=1\textwidth]{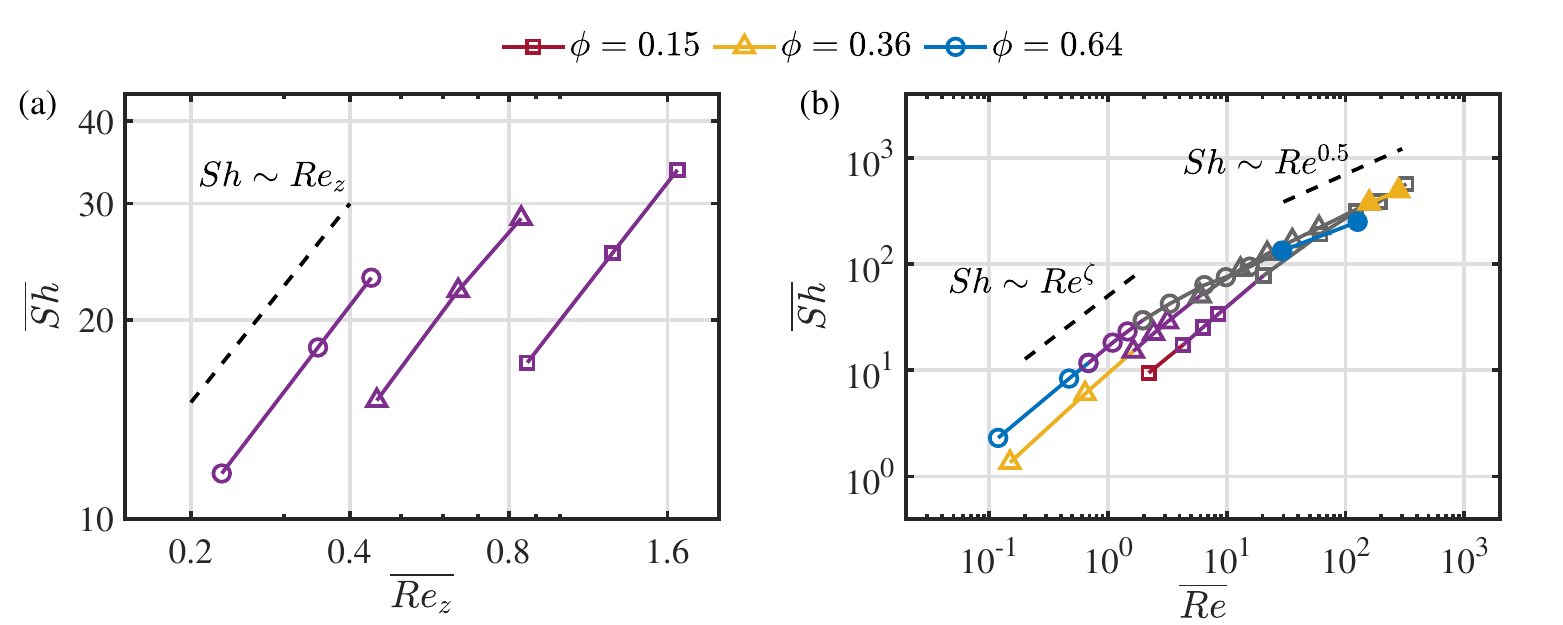}%
 \caption{The time-averaged Sherwood number $\overline{\She}$ versus (a) the time-averaged vertical Reynolds number $\overline{Re_z}$ calculated within the BL region and (b) the time-averaged Reynolds number $\overline{Re}$ calculated over the entire domain. The open coloured symbols denote the Darcy regime, the purple symbols denote the high-$\Ray_D$ Darcy regime, the grey symbols denote the transitional regime, and the solid symbols denote the RB regime. The exponent $\zeta$ in panel (b) is calculated by fitting the DNS data, which equals 0.90, 0.96 and 0.98 for $\phi=0.64$, 0.36 and 0.15, respectively.}
 \label{fig:GL-Re}
\end{figure}

Nevertheless, equation \eqref{eq:GL-Sh} does not directly provide the relationship between $\She$ and $\Ray$, as it also includes the Reynolds number $\Rey$. Therefore, we need to further derive a new equation based on concentration dissipation. For large $\Sch$, as $\Ray$ increases, the flow transitions from the so-called $\rom{1}_\infty$ regime to $\rom{3}_\infty$ regime \citep{grossmann2001thermal}, where $\epsilon_c$ shifts from being BL-dominated to bulk-dominated, consistent with our previous observations (figure \ref{fig:diss2}). In the high-$\Ray_D$ Darcy regime, $\epsilon_c$ is dominated by BLs, and the governing equation for concentration within BLs can be simplified from equation \eqref{eq:ns*b} as follows \citep{grossmann2000scaling}:
\begin{equation}
u_x^*\partial_x c^*+u_z^*\partial_z c^*=\kappa\partial_z^2 c^*. \label{eq:GL-BL}
\end{equation}
In the absence of porous media, the magnitudes of the two terms on the left side of the above equation are comparable. However, the presence of porous media changes the situation. Figure \ref{fig:GL-BLvelocity} shows the horizontal velocity field $u_x$ and the vertical velocity field $u_z$  near the boundary for a case with $\phi=0.64$ in the high-$\Ray_D$ Darcy regime. The concentration BL is also indicated by the yellow dashed lines in the figure. It can be seen that due to the presence of the porous media, there are many regions within the BL where $u_z$ is relatively large, slightly smaller than $u_x$, but of a comparable magnitude. This phenomenon is caused by the physical structure of porous media and is closely related to mechanical dispersion. In this situation, the second term on the left side of equation \eqref{eq:GL-BL} becomes more significant, and we can balance it with the right-side term. Considering $\partial_z \sim 1/\delta_c^*$, we can obtain:
\begin{equation}
\frac{u_z^*}{\delta_c^*}\sim \frac{\kappa}{\delta_c^{*2}} \label{eq:GL-BL2}
\end{equation}
If we define a vertical Reynolds number as $\Rey_z=\langle u^*_z \rangle^{rms}_{BL} H^*/\nu$, where $\langle \cdot \rangle^{rms}_{BL}$ denotes the root-mean-square value calculated over the concentration BL region (including the obstacles), and consider $\She=H^*/(2\delta_c^*)$, we can conclude from \eqref{eq:GL-BL2} that
\begin{equation}
\She \sim \Rey_z \Sch. \label{eq:GL-BL3}
\end{equation}
Figure \ref{fig:GL-Re}a illustrates the relationship between $\She$ and $\Rey_z$ in the high-$\Ray_D$ Darcy regime. By fitting the DNS data points, it can be observed that $\She$ and $\Rey_z$ indeed follow a linear scaling for various $\phi$.

If $\Rey_z \sim \Rey$, then by combining \eqref{eq:GL-Sh} and \eqref{eq:GL-BL3}, and considering $\She \gg 1$, we can obtain $\She \sim \Ray$, consistent with the classical theory \citep{howard1966convection}. However, the vertival velocity within BLs originates from the mechanical dispersion, thus $\Rey_z \sim \Rey$ does not always hold when the porous structure has an impact on it. In figure \ref{fig:GL-Re}b, we plot the relationship between $Sh$ and $Re$. By fitting the data points from the high-$Ra_D$ Darcy regime (represented by purple symbols), the corresponding $Sh \sim Re^\zeta$ scaling is obtained, with exponents of 0.90, 0.96, and 0.98 for the three porosities. This suggests a nonlinear scaling $Re_z \sim Re^\zeta$. In the RB regime, the data approaches a $Sh \sim Re^{0.5}$ scaling, which can be derived from the classical GL theory in the $\rom{3}_\infty$ regime \citep{grossmann2001thermal}. Substituting these two scaling laws into equation \eqref{eq:GL-Sh}, and considering $\She \gg 1$, we can obtain:
\begin{equation}\label{eq:GL-final}
  \She \sim \Ray^\gamma, \quad \gamma=\left\{  \begin{aligned}     
   &\frac{\zeta}{2-\zeta}  & \quad high-Ra_D \rm \quad Darcy \quad regime,\\
   & \quad \frac{1}{3}  &  \quad \rm RB \quad regime.
\end{aligned} \right. 
\end{equation}
Here, $\zeta$ is an empirical coefficient related to the mechanical dispersion effect near the boundary, reflecting the correlation between the vertical velocity within the BLs and the global velocity. As porosity decreases, $\zeta$ gradually approaches 1. Substituting its specific values into equation \eqref{eq:GL-final}, we obtain the $\gamma$ values for the three porosities as 0.82, 0.92, and 0.96, respectively, which is consistent with the previous results (see figure \ref{fig:regime}). Therefore, we successfully extend the GL theory to the RB region with obstacles and provid a physical explanation for the nonlinear scaling of mass transport in porous medium convection.

\section{Conclusion}\label{sec:conclusion}
In summary, we conducted a series of 2D pore-scale simulations to investigate density-driven convection in porous media. The porous structure was modeled by arranging obstacles in a regular pattern within the RB region, with porosities $\phi$ of 0.64, 0.36, and 0.15. With the Schmidt number set to 400, owing to the enhanced capabilities of the dual-resolution technique, we were able to set the values of Rayleigh-Darcy number $\Ray_D$ over a broad range, form $10^2$ up to $10^6$. As $\Ray_D$ increases, the flow regime sequentially transitions through the Darcy regime, transitional regime and RB regime. Correspondingly, the $Sh\sim Ra_D^\gamma$ relationship transitions from a sublinear scaling to the classic 0.3 scaling seen in RB convection. The exponent $\gamma$ in the high-$\Ray_D$ Darcy regime increases with decreasing $\phi$, i.e. $\gamma=0.81$, 0.92 and 0.97 for $\phi=0.64$, 0.36 and 0.15, respectively. A detailed analysis of the flow field structure revealed that when the plume width in the bulk region decreases to a scale comparable to the REV size $D$, the pore-scale structure starts to influence the flow, and the Darcy model starts to fail. The corresponding critical $Ra_D$ is approximately 4000 (for all porosities), which is still below the ultimate regime ($Ra_D > 10000$) found in the Darcy model. The flow finally enters the RB regime when the concentration BL thickness $\delta_c$ is less than $1/6$ of the pore space $l$, meaning that the plume width is much smaller than the spacing between obstacles. At this stage, the porous structure has minimal impact on mass transport.

In the Darcy regime, as $\Ray_D$ increases, the concentration field transitions from large-scale convection rolls to chaotic small-scale plumes. At this stage, the pore-scale Reynolds number remains much smaller than 1, indicating a ``pseudo-turbulent'' state. As $\Ray_D$ further increases, the flow enters the transitional regime, where large-scale periodic structures re-emerge, accompanied by fan-shaped structures caused by dispersion effect. When $\Ray_D$ becomes sufficiently high, the flow enters the RB regime, where dispersion effect diminishes, and the entire flow field consists of very small-scale plumes and two pairs of large-scale convection rolls. By calculating the horizontal auto-correlation coefficient of the concentration field, we can systematically examine the variation in the horizontal characteristic length scale of small-scale structures. In the Darcy regime, when the horizontal wavelength $\lambda_x$ is re-scaled by $D$, the cases with different $\phi$ collapse and the wavenumber approximately follows $k_x \sim \Ray_D^{0.4}$. In the transitional regime, $\lambda_x$ remains roughly constant at $2D$, indicating that individual plumes tend to remain within one REV. In the RB regime, although the porous structure has little effect on mass transport, $\lambda_x$ becomes proportional to $\delta_c$ and inversely proportional to $l$. This is primarily due to the horizontal dispersion of plumes after encountering obstacles. The above results indicate that in porous media convection, the REV size $D$ is a crucial intrinsic length scale. Similarly, when the vertical coordinate $z$ is re-scaled by $D$, the vertical profiles of concentration dissipation rate $\epsilon_c$ exhibit consistency at the same $Ra_D$. However, the kinetic energy dissipation rate $\epsilon_u$ shows a certain degree of lag, primarily due to the additional kinetic energy dissipation caused by the boundaries of internal obstacles.

Using GL theory, we successfully explained the changes in the scaling for $\She \sim \Ray_D^\gamma$ across different regimes. First, regarding the kinetic energy dissipation rate, the equation \eqref{eq:GL-Sh} that we previously derived for thermal convection \citep{liu2020rayleigh} also applies to the current cases across all the regimes, with only the empirical coefficient needing to be adjusted to 10. However, since this formula includes $Sh$, $Ra$ and $Re$, we still need to eliminate $Re$ based on the concentration dissipation rate. Using the discussion from $\rom{1}_\infty$ regime of GL theory, we demonstrated the linear relationship between $Sh$ and the vertical Reynolds number $Re_z$ within the BLs in the high-$\Ray_D$ Darcy regime. We further revealed the non-linear relationship between $Re_z$ and the global Reynolds number $Re$, with the underlying mechanism related to mechanical dispersion effect within the BLs. According to DNS results, $Sh$ and $Re$ follow a nonlinear scaling in the Darcy regime, with the exponent $\zeta$ approaching 1 as $\phi$ decreases. As $Ra$ increases and the flow enters the $\rom{3}_\infty$ regime, the GL theory indicates that $Sh \sim Re^{0.5}$, which is consistent with our DNS results. Based on the above analysis, we ultimately derived the scaling law \eqref{eq:GL-final} for $Sh$ and $Ra$ in both the high-$\Ray_D$ Darcy regime and the RB regime. Our findings shed some new light on understanding the nonlinear scaling that emerges in mass transport when considering pore-scale effects.

The starting point of our current work is geological CO$_2$ sequestration, where understanding the settling rate of CO$_2$ in saline aquifers is of significant practical importance. However, the current simulations still have some discrepancies with real-world conditions, including the non-random distribution of obstacles, the flow being confined to 2D and the fixed salinity distribution on both the upper and lower boundaries (where the real scenario is closer to one-sided convection). Despite these limitations, our current configuration provides a statistically steady state, allowing us to draw some meaningful conclusions. This is helpful for understanding convection in porous media itself, which can ultimately be applied to CO$_2$ sequestration. We are also working on the case where obstacles are randomly distributed in a 3D setting.

\backsection[Funding]{This work was supported by the National Natural Science Foundation of China under grant no. 11988102, the New Cornerstone Science Foundation through the New Cornerstone Investigator Program, the XPLORER PRIZE and the Postdoctoral Fellowship Program of CPSF under grant no. GZC20231219. Y.Y. also acknowledges the partial support from the Laoshan Laboratory Project under grant no. LSKJ202202000.}

\backsection[Declaration of interests]{The authors report no conflict of interest.}

\appendix
\section{The results with randomly distributed obstacles}\label{appA}
\begin{table} 
\begin{center}
\def~{\hphantom{0}}
\begin{tabular}{cccccccccccccccc}
  $\phi$ & $\Ray$ & $\Ray_D$ & $\Gamma$  & $N_x(m_x)$ & $N_z(m_z)$ & $t_d$ & $t_s$ & $\overline{\She}$ & $\overline{\Rey}$ & $\Delta_{Sh}(\%)$\\[5pt]
   0.64 & $1 \times 10^7$ & $2.6 \times 10^2$ &  2   &   1280(1)  &   640(1)  &  10000 &  3000  &  6.673  & 0.238 & 0.5 \\ 
   0.64 & $1 \times 10^8$ & $2.6 \times 10^3$ &  2   &   1280(1)  &   640(1)  &  5000 &  5000  &  24.06  & 1.391 & 0.4 \\ 
   0.64 & $1 \times 10^9$ & $2.6 \times 10^4$ &  2   &   1280(3)  &   640(3)  &  5000 &  2000  &  61.97  & 6.760 & 1.8 \\ 
   0.64 & $1 \times 10^{10}$ & $2.6 \times 10^5$ &  2   &   1280(4)  &   640(4)  &  5000 &  2000  &  120.3  & 29.92 & 1.1 \\ 
   0.64 & $1 \times 10^{11}$ & $2.6 \times 10^6$ &  2   &   1536(6)  &   768(6)  &  2500 &  1000  &  212.8  & 118.8 & 1.2 \\ 
\end{tabular}
\caption{Numerical details for the cases with randomly arranged obstacles. Columns from left to right are: the porosity, the Rayleigh number, the Rayleigh-Darcy number, the aspect ratio, the resolutions with refined factors in the horizontal and vertical directions, the simulation time before the statistical stage, the statistical time, the statistical Sherwood number, the statistical Reynolds number, and the relative difference of the statistical Sherwood numbers calculated at the two plates.  }
\label{tab:app}
\end{center}
\end{table}
\begin{figure}
 \centering
 \includegraphics[width=1\textwidth]{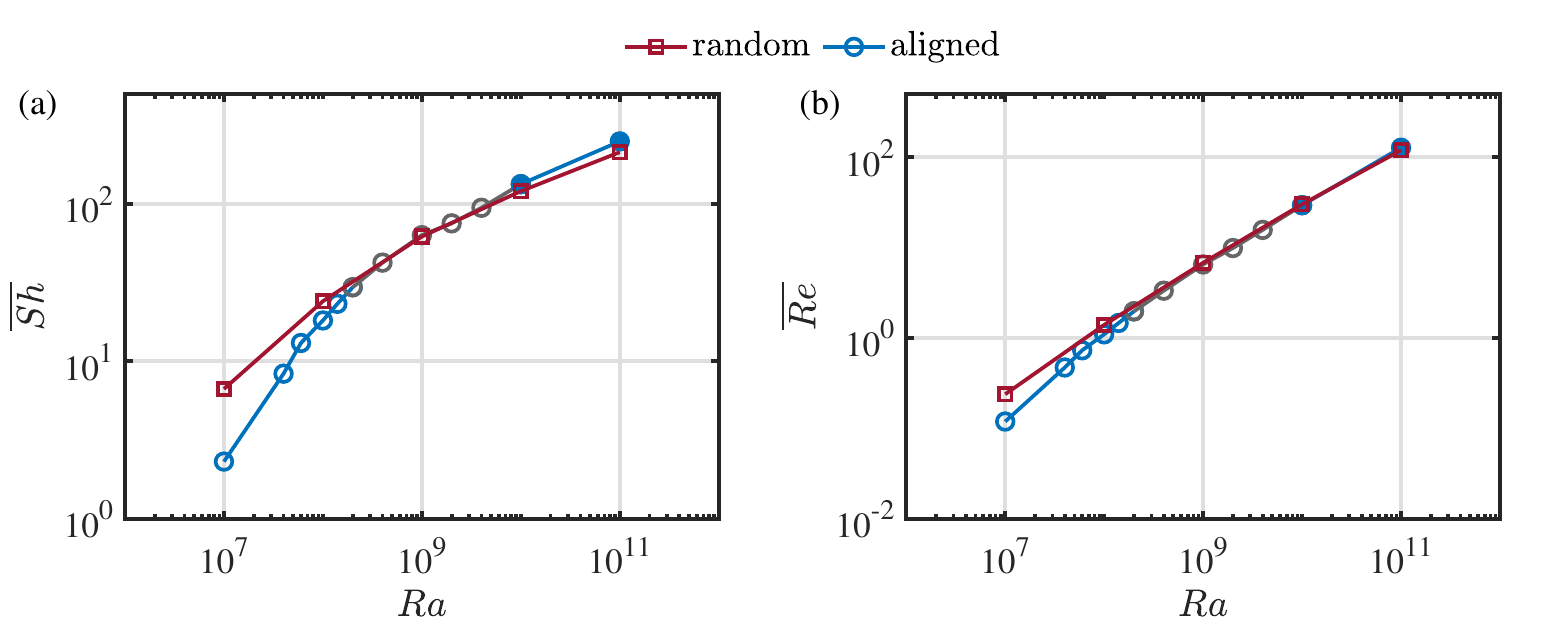}%
 \caption{(a) The time-averaged Sherwood number $\overline{\She}$ and (b) the time-averaged Reynolds number $\overline{Re}$ versus the Rayleigh number $\Ray$.}
 \label{fig:app1}
\end{figure}
\begin{figure}
 \centering
 \includegraphics[width=1\textwidth]{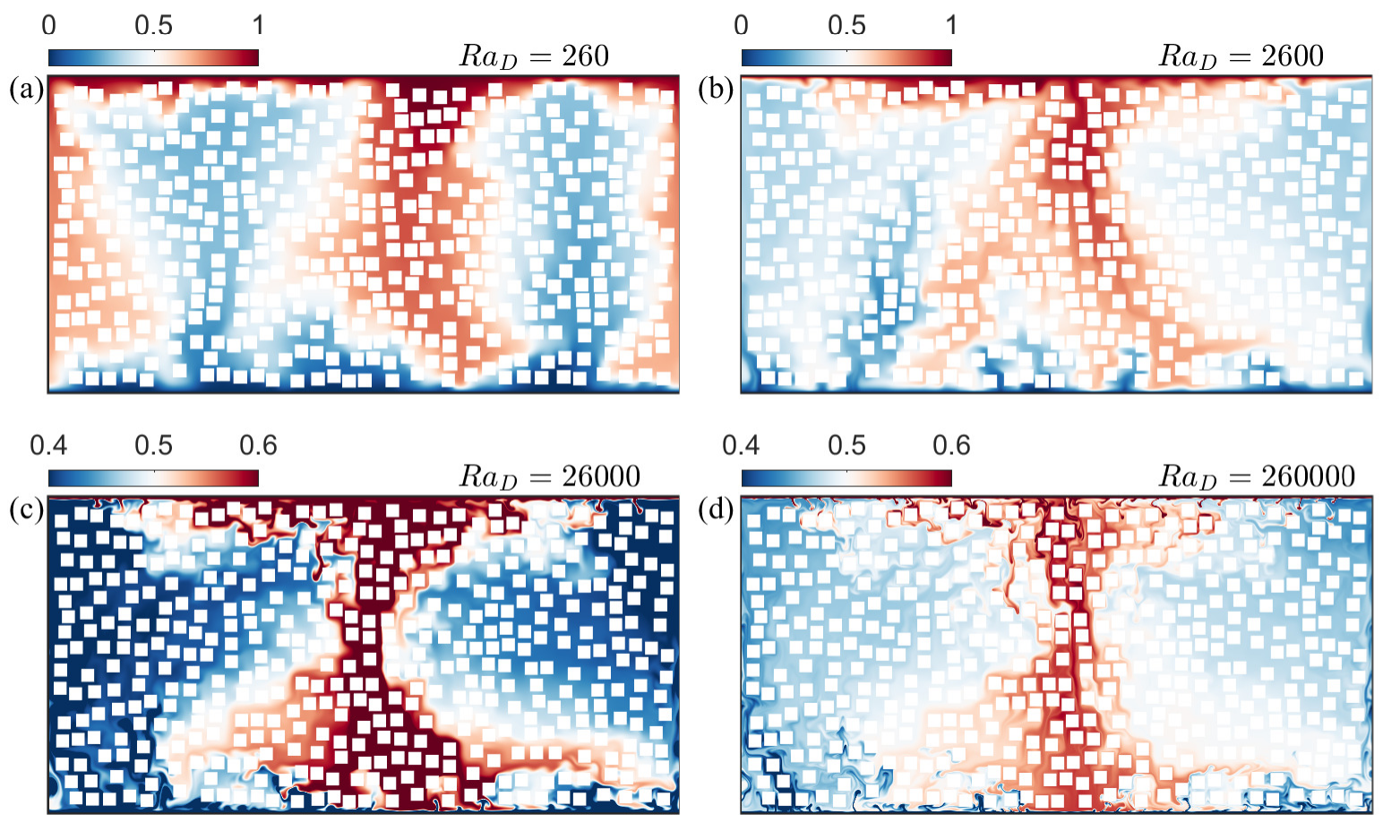}%
 \caption{Snapshots of the instantaneous concentration fields for the cases with randomly arranged obstacles. (a) $\Ray=10^7$, (b) $\Ray=10^8$, (c) $\Ray=10^9$ and (d) $\Ray=10^{10}$. The porosity is fixed at 0.64. The obstacles are denoted by the white squares.}
 \label{fig:app2}
\end{figure}
This appendix presents the DNS results with randomly distributed obstacles. We simulated five cases with  $\phi=0.64$ and $\Ray$ ranging from $10^7$ to $10^{11}$. The size of the obstacles remains $d=0.04$, and the minimum spacing between them is $l_{min}=0.008$. The resolution we set ensures that there are at least five grid points within $l_{min}$. Table \ref{tab:app} summarizes the numerical details of all cases. In figure \ref{fig:app1}, we plot variation of  $Sh$ and $Re$ with $Ra$ obtained from these cases, and we also include the results from the cases with aligned obstacles for comparison. It can be observed that the overall trends are consistent between the two. In the Darcy regime, $Sh$ and $Re$ of the cases with randomly distributed obstacles are slightly higher than those with aligned obstacles, but in the transitional and RB regimes, the results of both are nearly identical. For the scaling of $Sh \sim Ra^\gamma$ in the high-$\Ray_D$ Darcy regime, it can be predicted that the exponent $\gamma$ will be smaller for the randomly distributed obstacles. The differences in the flow field structures between the two distributions are more significant, as shown in figure \ref{fig:app2}. Compared to the flow field with aligned obstacles in figure \ref{fig:flowfield}, large-scale structures never disappear when obstacles are randomly distributed, and the dispersion effects are more prominent across all regimes. Furthermore, starting from the high-$\Ray_D$ Darcy regime, there is only one pair of large-scale convection rolls in the flow field. In summary, the distribution of obstacles does have an impact on mass transport efficiency and flow field structure. Specifically, the relationship between $\She$ and $\Ray$ further deviates from linearity in the high-$\Ray_D$ Darcy regime with the presence of randomly distributed obstacles. More systematic simulations and analyses will be conducted in the future.

\bibliographystyle{jfm}
\bibliography{porousref}


\end{document}